\documentclass{mn2e}
\usepackage{times,graphicx,astrobib,amssymb,amsmath,epsfig,subfigure,mathrsfs,epstopdf}
\pdfoutput=1
\newcommand{\de}{{\rm d}}
\newcommand{\bea}{\begin{eqnarray}}
\newcommand{\eea}{\end{eqnarray}}

\title[Cosmic ray driven outflows]
{Cosmic ray driven outflows from high redshift galaxies}
\author[Samui, Subramanian \& Srianand] 
{Saumyadip Samui\thanks{E-mail: samui@iucaa.ernet.in},
Kandaswamy Subramanian\thanks{E-mail: kandu@iucaa.ernet.in},
Raghunathan Srianand\thanks{E-mail: anand@iucaa.ernet.in} \\
IUCAA, Post Bag 4, Ganeshkhind, Pune 411 007, India.}

\begin{document}

\maketitle

\begin{abstract}

We study winds in high redshift galaxies driven by a relativistic
cosmic ray (proton) component in addition to the hot thermal gas
component. Cosmic rays (CRs) are likely to be efficiently generated
in supernova (SNe) shocks inside galaxies. We obtain solutions of
such CR driven free winds in a gravitational potential of the
Navarro-Frenk-White (NFW) form, relevant to galaxies. Cosmic rays
naturally provide the extra energy and/or momentum input to the
system, needed for a transonic wind solution in a gas with adiabatic
index $\gamma=5/3$. We show that cosmic rays can effectively drive
winds even when the thermal energy of the gas is lost due to radiative
cooling. These wind solutions predict an asymptotic wind speed closely
related to the circular velocity of the galaxy. Furthermore, the mass
outflow rate per unit star formation rate ($\eta_w$) is predicted to
be $\sim 0.2-0.5$ for massive galaxies, with masses
$M \sim 10^{11}-10^{12} M_\odot$. We show $\eta_w$ to be inversely
proportional to the square of the circular velocity. Magnetic fields
at the $\mu$G levels are also required in these galaxies to have a
significant mass loss. A large $\eta_w$ for small mass galaxies
implies that cosmic ray driven outflows could provide a strong
negative feedback to the star formation in dwarf galaxies. Further,
our results will also have important implications to the metal
enrichment of the intergalactic medium. These conclusions are
applicable to the class of free wind models where the source
region is confined to be within the sonic point.

\end{abstract}

\section{Introduction}

Cosmic rays are expected to be generated and accelerated in shocks
created by the exploding supernovae (SNe) in star forming regions.
The properties of cosmic rays (CRs) in our Galaxy are well documented
(Berezinskii et al., 1990; Schlickeiser, 2000; Dogiel, Schonfelder
\& Strong, 2002; Hillas, 2005; see also Kulsrud, 2005 for an excellent
introduction to the astrophysics of CRs). The energy density in the
proton component is about $1$~eV~cm$^{-3}$, and CRs are thought to
be confined to the Galactic disk for about $10^7$ yr before escaping,
presumably into the intergalactic medium (IGM). It is widely accepted
that $\ge 10\%$ of the average SNe energy must go into accelerating
the CRs so that the flux density of CRs can be maintained at the
observed value (Hillas, 2005).

Low energy CR protons play an important role in the ionization and
thermal state of photo-dissociation regions in the interstellar
medium of normal galaxies. As the direct interaction cross sections
are less for high energy cosmic rays, one would naively imagine that
they would hardly interact with matter. However CRs in our neighbourhood 
are observed to have a very small anisotropy (about 1 part in $10^4$).
This is explained by pitch-angle scattering of CRs by Alfv\'en waves
in our interstellar medium. These waves can in turn be generated
by the CRs themselves, by a streaming instability, as they gyrate
around and stream along the galactic magnetic field at a velocity
faster than the Alfv\'en velocity (Wentzel, 1968; Kulsrud \& Pearce, 1969;
Kulsrud \& Cesarsky, 1971). The waves travelling in the direction
of the streaming CRs (and satisfying a resonance condition) are
amplified by this streaming instability. The amplified waves have
a wavelength of order the cosmic ray gyro-radius, and then resonantly
scatter the CRs. Such scattering leads to transfer of momentum and
energy from CRs to the waves. This momentum and energy is in turn
transmitted to the thermal gas, when the waves damp out. If the CR
pressure is $P_c$, the force exerted per unit volume by the CRs on
the thermal gas is given by $-{\bf \nabla}P_c$, while the rate of 
energy transfer to the gas via the intermediary of the waves, is
given by $\vert{\boldsymbol v}_A\cdot{\bf \nabla}P_c\vert$
(Wentzel, 1971). Here ${\boldsymbol v}_A$ is the the Alfv\'en velocity
(to be defined below). The modulus is taken since CRs always lose
energy in generating the waves, while the gas always gains this energy
as the waves damp out. In addition, power is transferred to
the gas due to the work done by the CR pressure gradient
in the presence of a bulk velocity ${\boldsymbol v}$ of the fluid, and
this is given by $(-{\boldsymbol v}\cdot{\bf \nabla}P_c)$.

In our Galaxy the CR pressure support plays an important role in
the vertical force balance of the disk gas. It is also believed
that they may regulate star formation during the evolution of
galaxies (Ensslin et al. 2007; Socrates et al., 2008; Jubelgas
et al., 2008). The CR pressure gradient can also drive a wind
of thermal gas and CRs from the galaxy. The possibility that CRs
play an important role in driving outflows from our Galaxy has
been widely studied in the literature (Ipavich, 1975; Breitschwerdt
et al., 1987; 1991; 1993; Zirakashvili et al., 1996; Ptuskin et al., 1997; 
Breitschwerdt et al., 2002; Everett et al., 2008; 2009).
Such models can also explain better the observed diffuse soft X-ray
emission of our Galaxy (Everett et al., 2008). High redshift galaxies
are expected to have a much larger rate of star formation than our
Galaxy. Therefore, it is natural to consider the possibility of their
developing outflows driven by cosmic rays. This forms the main motivation
for the present work.

Furthermore, observations of high-$z$ Lyman break galaxies (LBGs) frequently
show evidence for galactic scale superwinds (Pettini et al., 2001).
Profiles of C~{\sc iv} and O~{\sc vi} absorption lines seen in the high
redshift damped Lyman-$\alpha$ systems (DLAs) are also consistent with
them originating from outflows in DLA protogalaxies (Fox et al., 2007a, 2007b).
The possible importance of outflows in the high redshift universe is
also indicated by the presence of metals in the IGM as traced by the
Lyman-$\alpha$ forest (Tytler et al., 1995; Songaila \& Cowie, 1996; 
Carswell et al., 2002; Simcoe et al., 2002;  Bergeron et al., 2002;
Rauch et al., 1997; Songaila, 2001; Ryan-Weber et al., 2006).
These metals can only originate in stars and have to be then transported
out into the IGM via outflows. Thus outflows appear to be ubiquitous
in high redshift galaxies.
 
There exist several semi-analytical and numerical models of outflows
aimed at explaining the presence of metals in the IGM (Nath \& Trentham,
1997; Efstathiou, 2000; Madau, Ferrara \& Rees, 2001; Furlanetto \& Loeb,
2003; Scannapieco, 2005; Samui, Subramanian \& Srianand, 2008). All
these models either use the pressure or momentum of the thermal gas,
generated from SNe explosions to drive the outflow. There have also
been models based on radiation pressure driving the winds
(Dijkstra \& Loeb, 2008; Nath \& Silk, 2009). The role played by
cosmic rays (which will be copiously produced in such high redshift
starburst galaxies) has not been explored so far in the 
context of high redshift galactic winds.

Moreover, it is well known that for a $\gamma=5/3$ thermal gas,
transonic wind solutions with the fluid velocity being accelerated
from subsonic to supersonic speeds exist only if there is 
some form of energy and/or momentum input till the critical point
(Lamers \& Cassinelli, 1999). An example of such a solution is that
due to Chevalier and Clegg (1985), where one assumes that mass and
thermal energy are injected into the flow upto the critical point.
The Chevalier and Clegg (1985) solution also ignores gravity,
assuming the gas is hot enough not to feel the gravitational
influence of the galaxy. It appears unnecessarily restrictive to
assume that such a situation of energy input till the sonic point
can always be obtained in a galaxy. On the other hand cosmic ray driving
can naturally provide the required energy and/or momentum input,
even if the energy sources (i.e. SNe) are well within the critical point,
or the thermal energy is lost due to radiative cooling.
This forms an additional motivation for considering such
cosmic ray driven winds from high redshift star forming galaxies.

In the present work, we construct models of CR driven winds from
star forming galaxies. Although our emphasis is on high redshift
galaxies, our wind solutions would be equally applicable to any 
galaxy where the star formation results in copious production of CRs.
The pioneering work by Ipavich (1975)
considered cosmic ray driven wind solutions in a point mass gravitational
potential. Here, we closely follow the methodology of Ipavich (1975).
However we take the gravitational potential to be described by the
standard NFW  profile (Navarro, Frenk \& White, 1997), relevant for galaxies.
The paper is organized as follows. In the next section we recall
the equations for steady, spherically symmetric, cosmic ray driven
outflows, and also derive the general wind equation. This allows us
to estimate the mass loss due to such outflows. Section~3 casts these
equations in dimensionless form suitable for numerical solution.
A suite of such solutions of cosmic ray driven outflows from high
redshift galaxies with masses ranging from $10^{8}-10^{12}~M_\odot$,
are presented in section~4. For the smaller mass galaxies, it turns
out that the gas may radiatively cool significantly, and so in
section~5 we consider a suite of cold, cosmic ray driven, wind solutions.
We end with a discussion of our results in section~6.

\section{The CR driven outflow}

Let us consider a steady spherically symmetric outflow of thermal 
gas of density $\rho$ and $\gamma=5/3$, and ultra relativistic cosmic
ray particles, with $\gamma = 4/3$. We assume that the cosmic ray
component can be described as a fluid with negligible mass density,
but having appreciable energy density $U_c$ and a pressure $P_c = U_c/3$. 
The spherically symmetric form of the fluid and cosmic ray equations
are given by (cf. Ipavich, 1975),
\begin{equation}
\rho v r^2 = q = {\rm constant},
\label{eqn_con_q}
\end{equation}
\begin{equation}
\rho v \frac{\de v}{\de r} = -\frac{\de P}{\de r} - \frac{\de P_c}{\de r} -
\frac{GM(r)\rho}{r^2},
\label{eqn_2}
\end{equation}
\begin{equation}
\frac{1}{r^2}\frac{\de}{\de r}\left[\rho v r^2 \left(\frac{1}{2}v^2 + 
\frac{5}{2}\frac{P}{\rho} \right)\right] = - \rho v r^2 \frac{GM(r)}{r^4} + I,
\label{eqn_3}
\end{equation}
\begin{equation}
\frac{1}{r^2}\frac{\de}{\de r}\left[4P_c r^2 \left(v + v_A\right)\right]=-I.
\label{eqn_4}
\end{equation}
Here $v$ and $P$ are the radial velocity and pressure of the thermal
gas, $v_A = B/(4\pi\rho)^{1/2}$ is the Alfv\'en velocity, with $B$
describing the radial component of the magnetic field. As in Ipavich (1975),
we assume the magnetic flux, given by $B(r) (4\pi r^2)$ to be constant.
The equation of continuity describing mass conservation, is integrated
to give Eq.~\ref{eqn_con_q}. The momentum equation (Eq.~\ref{eqn_2}),
includes pressure gradients due to both the thermal and cosmic ray
components, and the gravitational attraction due to a mass distribution $M(r)$.
Energy conservation is described by Eq.~\ref{eqn_3}, where the $I$ term,
\begin{equation}
I = -\left( v + v_A \right) \frac{\de P_c}{\de r}
\label{eqn_I}
\end{equation}
incorporates the exchange of energy between the cosmic ray and thermal gas
components. The cosmic ray evolution itself is described by Eq.~\ref{eqn_4}.

We can also add Eqs.~\ref{eqn_3} and \ref{eqn_4} describing respectively
energy conservation and CR evolution, and integrate the resulting equation to
get another algebraic conservation law,
\begin{equation}
\frac{1}{2}q v^2 + \frac{5}{2}P v r^2 - \frac{qGM}{F(c)r}\ln\left(1+
\frac{cr}{r_{vir}}\right) + 4P_c r^2 \left(v + v_A\right) = C_e
\label{eqn_18}
\end{equation}
where $C_e$ is some constant of integration. If we multiply the above
equation by $4\pi$, then the terms on the LHS are respectively the
rate at which, the kinetic energy, enthalpy of the thermal gas,
gravitational energy and cosmic ray `enthalpy' are flowing through
a sphere of radius $r$. The sum is a constant at any radius and equal
to $4\pi C_e$. In the inner regions of the wind, the kinetic energy
contribution will be small, and the thermal and CR pressures contributions
will have to be sufficiently large to overcome the gravity. As $r\to \infty$,
all terms except the kinetic energy contribution will go to zero,
and $C_e \to q (v^2_\infty/2)$, where $v_\infty$ is the wind
velocity as $r\to \infty$.

A straightforward manipulation of the above equations, which we describe
in Appendix~\ref{appA}, allows us to derive the wind equation,
\begin{equation}
\frac{\de v}{\de r} = \cfrac{2v}{r}\cfrac{\left[1-\cfrac{GM(r)}{2rc_*^2}\right]}
{\left[\cfrac{v^2}{c_*^2}-1\right]},
\label{eqn_dv_dr}
\end{equation}
where the effective sound speed $c_*$ is determined by,
\begin{equation}
c_*^2 = \frac{5}{3}\frac{P}{\rho} + \frac{4}{3} 
\frac{P_c}{\rho}
\frac{(v + v_A/2)(3v-2v_A)}{3v(v+v_A)}.
\label{eqn_c_star}
\end{equation}
One sees that flow has to pass through a critical point where both 
the numerator and denominator of Eq.~\ref{eqn_dv_dr} has to vanish.
Indeed the numerator has to go to zero faster than the denominator,
to obtain a regular solution for the flow. At the critical point,
$r=l$, the velocity of the gas is given by,
\begin{equation}
v(l) = c_*(l) = \sqrt{GM(l)/2l} = V_c/\sqrt{2},
\end{equation}
where we have defined the circular velocity of the galactic 
potential at the critical point $r=l$ to be  $V_c = (GM(l)/l)^{1/2}$. 
Note that $V_c$ is also roughly the circular velocity at the virial radius 
$r=r_{vir}$, if the galaxy has approximately a flat rotation curve;
that is $V_c \sim v_{cir} = (GM(r_{vir})/r_{vir})^{1/2}$.

We can now estimate the mass outflow rate as follows. Let the total
luminosity that is being pumped by the SNe to the outflowing material
be $L_0$, and assume that only a fraction $f$ of this ends up as the
wind kinetic energy. This fraction depends on the gravitational potential
of the galaxy and the distribution of mass and energy sources with respect
to that potential. In Appendix~\ref{appB} we have given an estimate of $f$
for a uniform source distribution and find a typical value of $f\sim 0.1$.
Moreover by solving the fluid equations numerically, we will generally find that
the asymptotic velocity of the outflowing gas as $r\to \infty$, $v_\infty$, 
is closely related to the velocity at critical point and hence the circular
velocity, i.e.
\begin{equation}
v_\infty = F v(l) = \frac{F}{\sqrt{2}} V_c,
\end{equation}
with $F/\sqrt{2}$ is of order unity.
Then,
\begin{equation}
4\pi C_e = \frac{1}{2}{\dot M}_w v_\infty^2 = fL_0 =
f (\nu {\dot M}_{SF})(\epsilon_w E_{SNe}).
\label{eqn_M_dot_w_L0}
\end{equation}
Here ${\dot M}_w = 4\pi q$ is the mass outflow rate, ${\dot M}_{SF}$
is the star formation rate, $\nu$ is the number of SNe produced per
unit mass of stars formed, $E_{SNe}=10^{51}$~erg is the energy produced
per SNe and $\epsilon_w$ is the fraction of that energy that drives the
outflow \footnote{A qualitatively similar estimate, without the factor $f$,
is given by Murray, Quataert \& Thompson (2005).}.
Hence the mass outflow rate is given by,
\begin{equation}
{\dot M}_w = 4\pi q = \frac{4 fL_0}{F^2 V_c^2}=
\frac{4 f \nu {\dot M}_{SF}\epsilon_w E_{SNe}}
{F^2 V_c^2}.
\label{eqn_M_dot_w_q}
\end{equation}
The factor $\nu$ depends on the initial mass function of the stars formed,
and is $\sim 1/130~M_\odot^{-1}$ for a Salpeter IMF with a lower and upper
cut-off masses of $m_{low} = 0.1~M_\odot$ and $m_{up}= 100~M_\odot$. 
Changing $m_{low}$ to $1~M_\odot$ gives $\nu = 1/50~M_\odot^{-1}$. 
We will adopt $\epsilon_w \sim 0.1$. The wind mass loading factor is then,
\begin{equation}
\eta_w = \frac{{\dot M}_w}{{\dot M}_{SF}} = \frac{4f}{F^2}
\left(\frac{\epsilon_w}{0.1}\right)\left(\frac{220~{\rm km~s}^{-1}}{V_c}
\right)^2 \left(\frac{100~M_\odot}{\nu^{-1}}\right),
\label{eqn_eta}
\end{equation}
where for the numerical estimate, we have scaled by some typical
values of $\epsilon_w$, $\nu$ and $V_c$. Our estimate in Eq.~\ref{eqn_eta}
leads to one of the important results of our work, that the mass loading
factor is expected to increase with a decreasing galaxy circular velocity,
approximately as $\eta_w \propto V_c^{-2} \propto v_{cir}^{-2}$.
For a massive galaxy with $V_c$ like our Galaxy, and taking $f \sim 0.1$,
$F \sim \sqrt{2}$ (see below), we expect to get $\eta_w \sim 0.2$.
On the other hand CR driven outflows from dwarf galaxies with much
smaller $V_c$ are expected to lead to a much higher mass loading factor.
The numerical solutions which are obtained below, bear out these general
expectations. 

\section{Numerical Solutions}

We now consider numerical solutions of the  fluid equations including
the CR component. Note that during most of its evolution the outflow
traverses through the dark matter halo of a galaxy, whose potential is
generally well described by the standard NFW form (Navarro, Frenk \& White, 1997).
The mass distribution associated with the NFW potential is given by,
\begin{equation}
M(r) = \frac{M}{{\mathfrak{F}}(c)}{\mathfrak{F}}\left(\frac{cr}{r_{vir}}\right)
\label{eqn_NFW}
\end{equation}
where the function ${\mathfrak{F}}(z)$ is given by,
\begin{equation}
{\mathfrak{F}}(z) = \ln (1 + z) - \frac{z}{ 1 + z }.
\label{eqn_F_c}
\end{equation}
Here $M$ is the total galaxy mass, $r_{vir}$ the virial radius of the halo,
and $c$ the concentration parameter, which decides the radius beyond which
the rotation velocity transits to an almost flat form. We choose $c$ to be
larger than what would obtain for just the dark halo, in order to mimic
the fact the observed rotation curves of disk galaxies are flat well within
their optical radius. Therefore, our model NFW potentials provide a reasonable
approximation to the gravitational potential of the galaxy, at least in
the spherical limit.

In order to obtain the wind solutions,we first transform all the equations to
a dimensionless form. We use the location of the critical point, say $l$, as
the length variable and define a dimensionless radius $x=r/l$. Thus the critical point
occurs at $x=1$. Using the constants $q$, $C_e$ and $l$ we define
the dimensionless density ($\rho^*$), velocity ($u$), thermal pressure ($\theta_P$)
and cosmic ray pressure ($\theta_{\Pi}$) respectively as
\begin{eqnarray}
\rho^* &=& 2 \rho \frac{l^2}{q}\left(\frac{2C_e}{q}\right)^{1/2},
\qquad u = v \left( \frac{q}{2C_e}\right)^{1/2},
\nonumber \\
\theta_P &=& P\frac{l^2}{C_e}\left(\frac{2C_e}{q}\right)^{1/2},
\qquad 
\theta_{\Pi} = P_c \frac{l^2}{C_e}\left(\frac{2C_e}{q}\right)^{1/2}.
\label{transform}
\end{eqnarray}

In terms of these dimensionless variables, the evolution equations
(\ref{eqn_con_q}), (\ref{eqn_2}), (\ref{eqn_18}) and (\ref{eqn_3})
respectively, for the CR driven outflow become
\begin{equation}
\rho^* u x^2 = 2,
\label{eqn_21}
\end{equation}
\begin{equation}
2\frac{\de u}{\de x} + x^2 \frac{\de \theta_P}{\de x} + \frac{4a}{ux^2}
\left[\ln (1+c' x) - \frac{c' x}{1+c' x}\right] + 
x^2\frac{\de \theta_\Pi}{\de x} = 0,
\label{eqn_22}
\end{equation}
\begin{equation}
u^2 + \frac{5}{2} \theta_P u x^2 - \frac{4a}{x} \ln (1+c'x) + 4\theta_\Pi x^2
\left[ u + \frac{(ub)^{1/2}}{x}\right] = 1,
\label{eqn_23}
\end{equation}
\begin{eqnarray}
\frac{\de }{\de x}\left[u^2 + \frac{5}{2}u x^2 \theta_P\right.
& - & \left.\frac{4a}{x}\ln(1+c' x) \right] \nonumber \\
& = & -x^2\left[ u + \frac{(ub)^{1/2}}{x}\right] \frac{\de \theta_\Pi}{\de x}.
\label{eqn_24}
\end{eqnarray}
Here we have also defined the dimensionless parameters of the system,
\begin{equation}
a = \frac{GM}{2l{\mathfrak{F}}(c)}\left(\frac{q}{2C_e}\right),
\label{eqna}
\end{equation}
\begin{equation}
b = \frac{v_A^2}{v}\frac{r^2}{l^2}\left(\frac{q}{2C_e}\right)^{1/2}
= \frac{(Br^2)^2}{4\pi q} \frac{1}{l^2} \left(\frac{q}{2C_e}\right)^{1/2},
\label{eqnb}
\end{equation}
\begin{equation}
c' = c \frac{l}{r_{vir}}.
\label{eqncprime}
\end{equation}
The parameter `$a$' represents the ratio of a typical gravitational
potential energy to thermal plus CR energy, while the parameter $b$
gives a measure of the strength of the magnetic field. These definitions
are similar to that adopted by Ipavich (1975). The parameter $c'$
gives the dimensionless measure of the location of the critical point,
in terms of the scale length $r_{vir}/c$ associated with a dark matter
halo. This parameter was not present in the work of Ipavich (1975),
as there gravity is due to a point mass. Its presence increases the
complexity of finding transonic solutions. Note that we need to solve
only two differential equations, due to the availability of the two
algebraic constraints representing mass and energy conservations.
We first solve for gas pressure, using Eq.~\ref{eqn_23}:
\begin{eqnarray}
\theta_P & = & \frac{2}{5 u x^2} \left[1 + \frac{4a}{x}\ln(1+c' x)\right.
\nonumber \\
& & \hskip 1.3cm - \left. u^2
- 4 \theta_\Pi x^2 \left\{ u + \frac{(ub)^{1/2}}{x}\right\} \right].
\label{eqn_25}
\end{eqnarray}
Substituting this in Eq.~\ref{eqn_24}, we obtain
\begin{equation}
\frac{\de \theta_\Pi}{\de x} = \cfrac{- \cfrac{4}{3}\theta_\Pi\left[\cfrac{2}{x}
+ \cfrac{1}{u}\cfrac{\de u}{\de x}\right]\left[u + \cfrac{(ub)^{1/2}}{2x} \right]}
{\left[ u + \cfrac{(ub)^{1/2}}{x}\right]}.
\label{eqn_26}
\end{equation}
Next we substitute Eq.~\ref{eqn_25} and Eq.~\ref{eqn_26} into
Eq.~\ref{eqn_22} to get the velocity gradient:
\begin{equation}
\frac{\de u}{\de x} = \cfrac{2u}{x} \cfrac{ \left[ \cfrac{a}{x}\ln(1+c' x)
+ \cfrac{3 a c'} {(1 + c' x)} - u^2 + 1 -{\cal F} \right] }{\left[ 4u^2 -
 \cfrac{4a}{x}\ln(1+c' x) -1 + {\cal F} \right] }
\label{eqn_27}
\end{equation}
where
\begin{equation}
{\cal F} = \cfrac{\theta_\Pi x^2}{3\left[ u + \cfrac{(ub)^{1/2}}{x}\right]}
\left[ 6 u^2 + 25 \cfrac{u}{x} (ub)^{1/2} + 14 \cfrac{ub}{x^2}\right].
\label{eqn_29}
\end{equation}
There is again a critical point to the solution where both
the numerator and denomination in Eq.~\ref{eqn_27} vanish.
These conditions imply that at the critical point, where $x=1$,
the velocity $u =u_c$, where
\begin{equation}
u_c^2 = a\left[\ln \left(1 + c'\right) - \frac{c'}{1+c'}\right] = a {\mathfrak{F}}(c')
\label{eqn_41}
\end{equation}
and
\begin{equation}
{\cal F} = 1 + \frac{4ac'}{1+c'}.
\label{eqn_42}
\end{equation}
The cosmic ray pressure at the critical point, $\theta_\Pi^c$,
can be obtained using Eq.~\ref{eqn_29} and \ref{eqn_42}, 
\begin{equation}
\theta_\Pi^c = \left( 1 + \frac{4ac'}{1+c'}\right)
\frac{\left[3\left\{ u_c + (u_cb)^{1/2}\right\}\right] }
{\left[ 6 u_c^2 + 25 u_c(u_cb)^{1/2}+ 14 u_cb\right]}.
\label{eqn_43}
\end{equation}
Since both numerator and denominator vanish at the critical point,
we use L'Hospital rule to evaluate the velocity gradient
(i.e. Eq.~\ref{eqn_27}) there. This results in a 
quadratic equation for $du/dx$ at the critical point given by,
\begin{eqnarray}
\left( 8u - {\cal B}\right) \left(\frac{\de u}{\de x}\right)^2 & + & 
\left(8 u^2 - 2u{\cal B} -{\cal A}\right)\frac{\de u}{\de x} \nonumber \\
& + & 2u \left( u^2 -{\cal A} \right)
+ \frac{6uac'^2}{\left(1+c'\right)^2} = 0
\label{eqn_44}
\end{eqnarray}
where
\begin{equation}
{\cal A}=\frac{\theta_\Pi\left[12u^3 + 95u^2(ub)^{1/2} + 62 u^2 b + 
14ub(ub)^{1/2}\right]}{9\left[u + (ub)^{1/2}\right]^2},
\label{eqn_A}
\end{equation}
\begin{equation}
{\cal B}=\frac{\theta_\Pi\left[12u^2 + 95u(ub)^{1/2} + 62 u b + 
14b(ub)^{1/2}\right]}{18\left[u + (ub)^{1/2}\right]^2}
\label{eqn_B}
\end{equation}
with all quantities in the above three equations evaluated at $x=1$
and $u=u_c$. Note that ${\cal A} = 2 u_c {\cal B}$.

Our numerical solutions start from the critical point which is at $x=1$.
At the critical point both the numerator and denominator of Eq.~\ref{eqn_27}
vanish though $\de u/\de x$ is finite. Hence, we calculate $\de u/\de x$
at the critical point from Eq.~\ref{eqn_44}, after adopting specific values
for the parameters $a$, $b$ and $c'$. We take the positive root of this
equation since we want a subsonic flow to transit to a supersonic flow.
We go both forward and backward in $x$ and solve the coupled differential
equations, Eq.~\ref{eqn_26} and Eq.~\ref{eqn_27}. This gives the dimensionless
velocity $u(x)$ and the dimensionless CR pressure $\theta_\Pi(x)$.
Then we use the algebraic Eqs.~\ref{eqn_21} and \ref{eqn_25} 
to obtain the dimensionless $\rho_*(x)$ and $\theta_P(x)$, respectively.
These dimensionless quantities can be converted to the dimensional
quantities ($\rho,v,P,P_C$), using the transformations of Eq.~\ref{transform}.

\section{Results}

The numerically computed CR wind solutions for several different
galactic masses, $10^8~M_\odot$ to $10^{12}~M_\odot$, and different
parameters, are given below. Table~\ref{table1} summarizes the assumed
parameters and the derived physical quantities associated with each
of these solutions. The double solid line in the table separates
the assumed parameters (above this line) and the derived quantities
(below the line).
\begin{figure*}
\centerline{ \includegraphics[width=12.0cm,angle=-90]{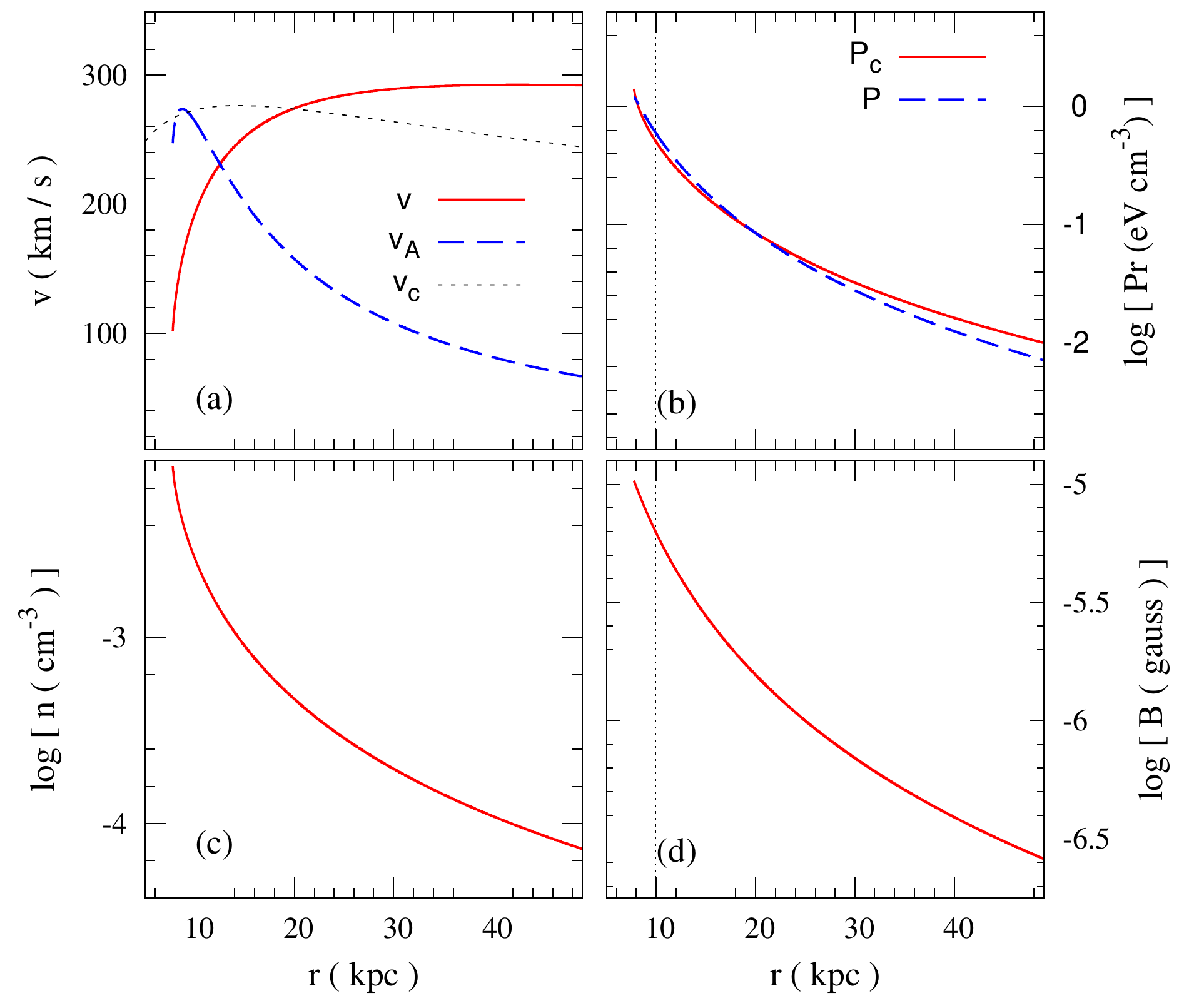}}
\caption[]{Characteristics of cosmic ray driven wind coming from a
$10^{12}~M_\odot$ galaxy. We show velocity, thermal gas pressure, cosmic ray
pressure, the density of the thermal gas and the magnetic field as a
function of radial distance from the galactic centre. The vertical dotted
lines in each panel represent the critical point. We also show the circular
velocity of the galaxy by dotted line in panel~(a).
}
\label{fig_wind_M12}
\end{figure*}
\subsection{CR driven wind from a massive galaxy}

First consider the example of wind evolution in the potential 
of a $10^{12}~M_\odot$ galaxy described by the NFW density profile.
We model the parameters of this potential to mimic outflows from a
massive high redshift galaxy, by adopting $c=15$ and a virial radius
$r_{vir} = 100$ kpc. We take the critical point to occur at
$l=r_{vir}/10$ or $l=10$ kpc. This fixes the value of $c'=1.5$.
We also take $a=2.0$ and $b=1.5$ and solve the dimensionless wind
evolution equations. 

In order to convert the wind solution to a dimensional
form, we also have to adopt a value for $C_e$.
Recall that $4\pi C_e = fL_0$, with $f \sim 0.1$ (Eq.~\ref{eqn_M_dot_w_L0}
and Appendix~\ref{appB}). Here, as mentioned earlier, $L_0$ is the rate of
energy input by SNe in the galaxy, and is given by,
\begin{equation}
L_0= 3 \times 10^{40} \left(\frac{{\dot M}_{SF}}{1 M_\odot {\rm yr}^{-1}}\right)
\left(\frac{\epsilon_w}{0.1}\right)\left(\frac{100~M_\odot}{\nu^{-1}}\right)
{\rm erg}~ {\rm s}^{-1}.
\label{energyL0}
\end{equation}
For the solution shown in Fig.~\ref{fig_wind_M12}, we have assumed a
star formation rate (SFR) ${\dot M}_{SF} \sim (10/f)~M_\odot {\rm yr}^{-1}$,
which leads to $4\pi C_e= 3 \times 10^{41}$~erg~s$^{-1}$.
This SFR is typical of that obtained in massive high redshift
Lyman break galaxies (Erb et al., 2006).
The assumed and derived parameters are summarized as model~A in
Table~\ref{table1}.

The fluid velocity obtained for this model is shown in panel~(a)
of Fig.~\ref{fig_wind_M12} as a solid line. The wind starts from a 
finite subsonic velocity ($v < c_*$) at $r_0=7.8$~kpc. It passes
through the critical point at $r=10$~kpc and the velocity continues
to increase before reaching an asymptotic velocity $\sim 292$~km~s$^{-1}$.
This should be compared with the circular velocity at the critical point,
$v_c \sim 270$~km~s$^{-1}$ and thus for this solution, $F/\sqrt{2} \sim 1.07$.
Panel (a) of Fig.~\ref{fig_wind_M12} also shows the circular velocity
of the galaxy (as a dotted line), and the Alfv\'en velocity ($v_A$)
(as a dashed line). The circular velocity is reasonably constant,
varying by about $10\%$ between $5$ and $50$ kpc. From the figure
it is also clear that the Alfv\'en velocity decreases with radius
after a small initial increase (like in the solutions given by Ipavich (1975)).
The cosmic ray pressure and the thermal pressure are shown by solid
and dashed line respectively in panel~(b) of Fig.~\ref{fig_wind_M12}.
At the critical point $P_c \sim 0.50$~eV~cm$^{-3}$ and $P \sim 0.59$~eV~cm$^{-3}$.
In the present solution, the CR pressure is comparable to the thermal
gas pressure at the base of the flow, then goes below the thermal
pressure and then again becomes higher at large radii. This latter
feature arises because $\gamma=4/3$ for the cosmic ray component
whereas $\gamma=5/3$ for the thermal gas. As the wind expands out,
the thermal gas pressure then decreases faster with radius compared
to the cosmic ray pressure.

In panel~(c) of Fig.~\ref{fig_wind_M12} we show the density
of the wind material. At the base of the outflow the density
is $\sim 0.008$~cm$^{-3}$ and falls of asymptotically as
$r^{-2}$, once the velocity of the flow becomes nearly
constant. Note that if we have a mass of $10^{11} M_\odot$
of baryons in a galaxy distributed within $10$ kpc, the average
density would be 1 cm$^{-3}$. Thus the gas density
required in the wind solution is modest in comparison.
The resulting temperature at the critical point is
$2.6 \times 10^6$~K and falls of monotonically with radius thereafter.

Finally, we show the magnetic field strength in the wind material in
panel~(d) of Fig.~\ref{fig_wind_M12}. Recall that we assumed
$B(r) \propto r^{-2}$. The proportionality constant is fixed from
the parameter $b$ of the model. We find the value of $B$ at the
critical point to be $\sim 6.2 \mu$G. This is comparable to typical
values of a few to $10\mu$G, obtained for the mean field in nearby
spiral galaxies (Beck, 2008), whereas it is smaller than the total
fields of $50-100 \mu$G inferred in some nearby starburst galaxies
(Chyzy \& Beck, 2004). We find that a threshold minimum value of the
magnetic field is required to obtain a cosmic ray driven wind passing
through the critical point. If the field were smaller than this threshold, 
the cosmic ray pressure becomes negative at some point. For the
dimensionless parameters $a=2.0$, $c'=1.5$ that we have adopted,
we require the parameter determining the magnetic field strength
$b \gtrsim 0.5$ to obtain a transonic wind solution. Thus one still
requires $\mu$G fields to have CR driven transonic winds. 
This requires efficient dynamo generation mechanisms to
operate in high redshift galaxies (cf. Brandenburg \&
Subramanian, 2005; Shukurov, 2007; Sur, Shukurov \& Subramanian, 2007).
Recent observations of Faraday rotation towards quasars showing Mg II
absorption at high redshift $z \sim 1-2$, suggests that organized
magnetic fields of high strengths $\sim 10\mu$ G, could be associated
with normal galaxies even at these redshifts (Bernet et al., 2008;
Kronberg et al., 2008).

\begin{figure*}
\centerline{
\subfigure[]{\includegraphics[width=5.5cm,angle=-90]{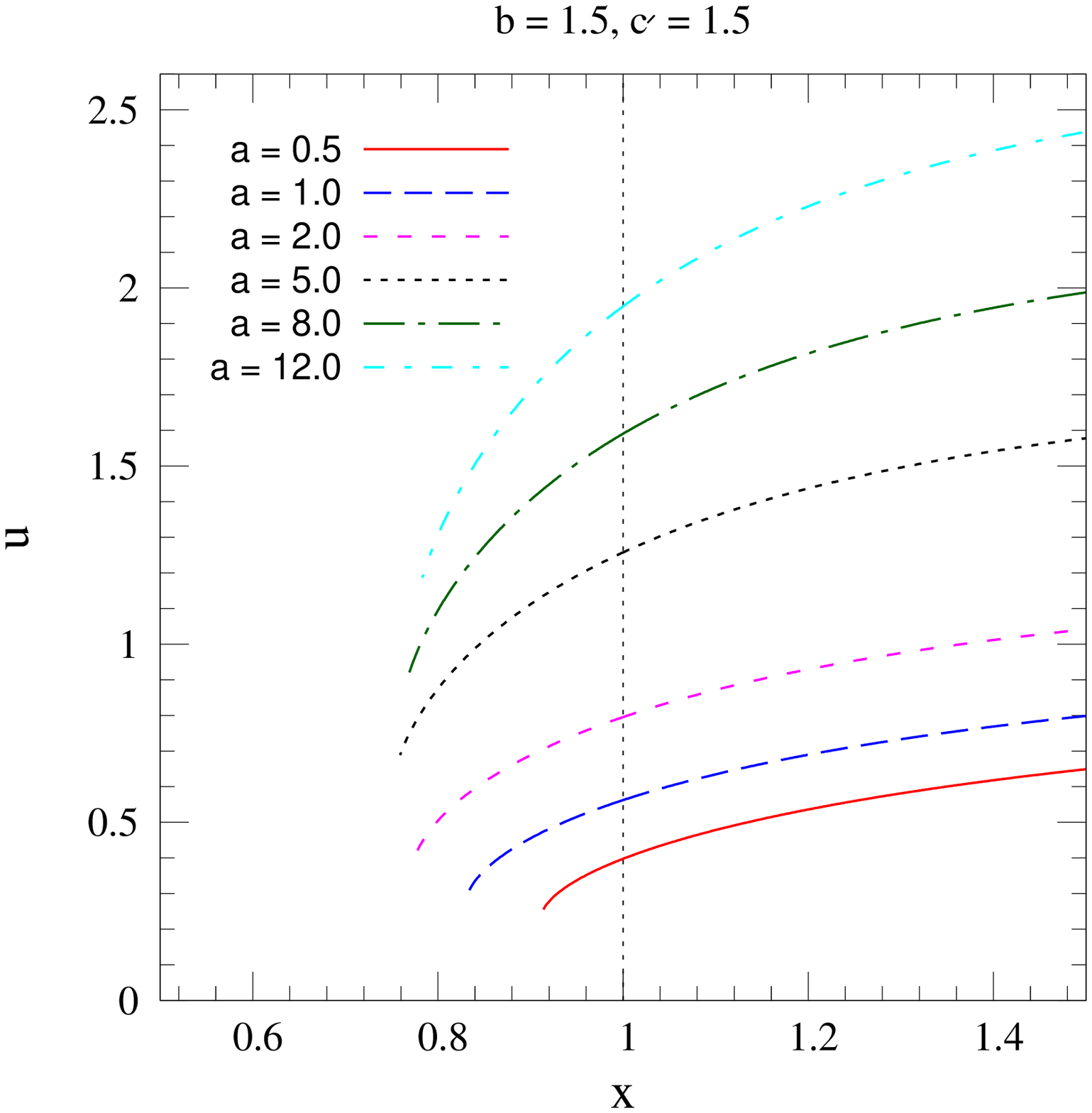}}
\label{fig_var_a}
\subfigure[]{\includegraphics[width=5.5cm,angle=-90]{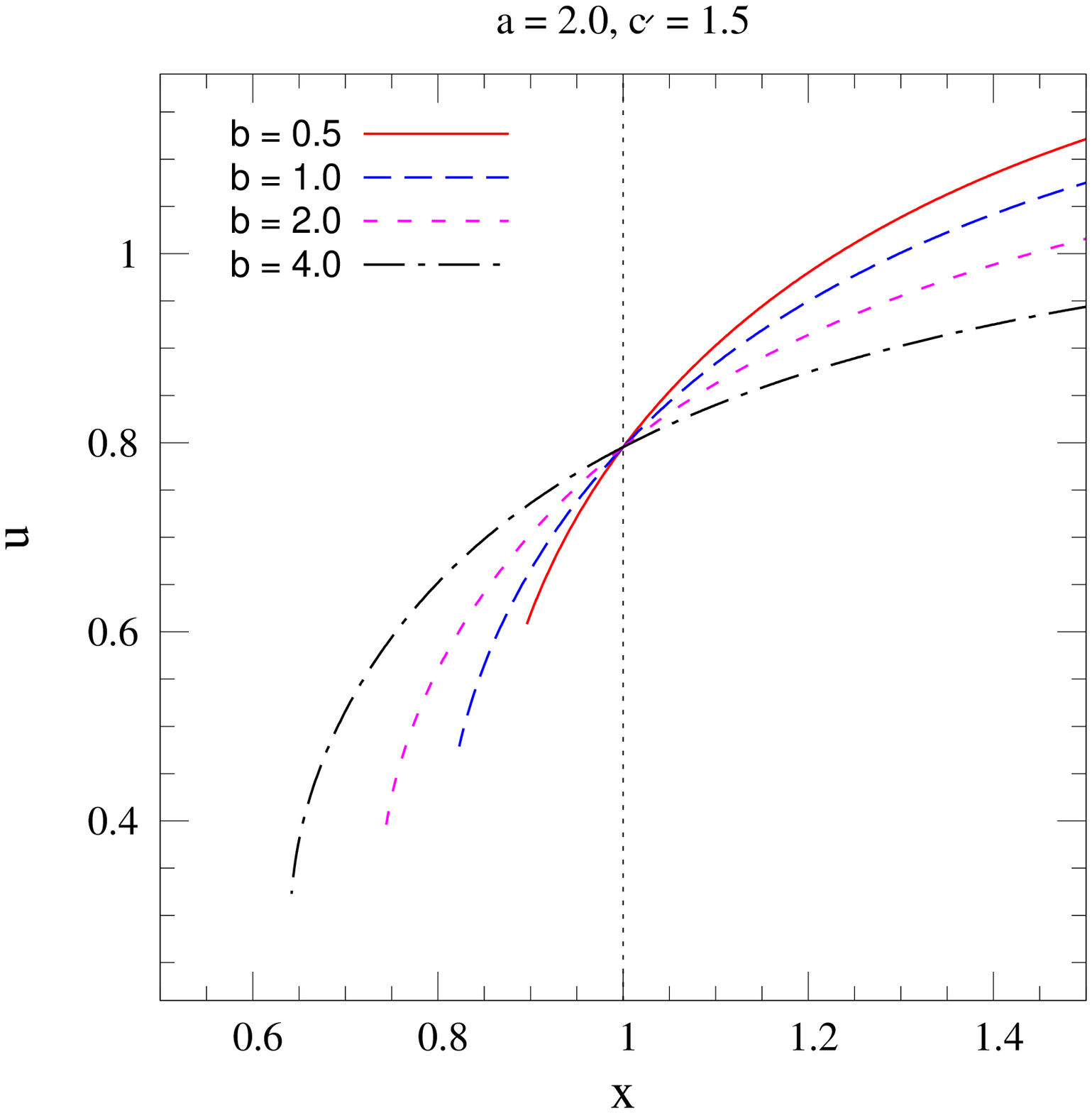}}
\label{fig_var_b}
\subfigure[]{\includegraphics[width=5.5cm,angle=-90]{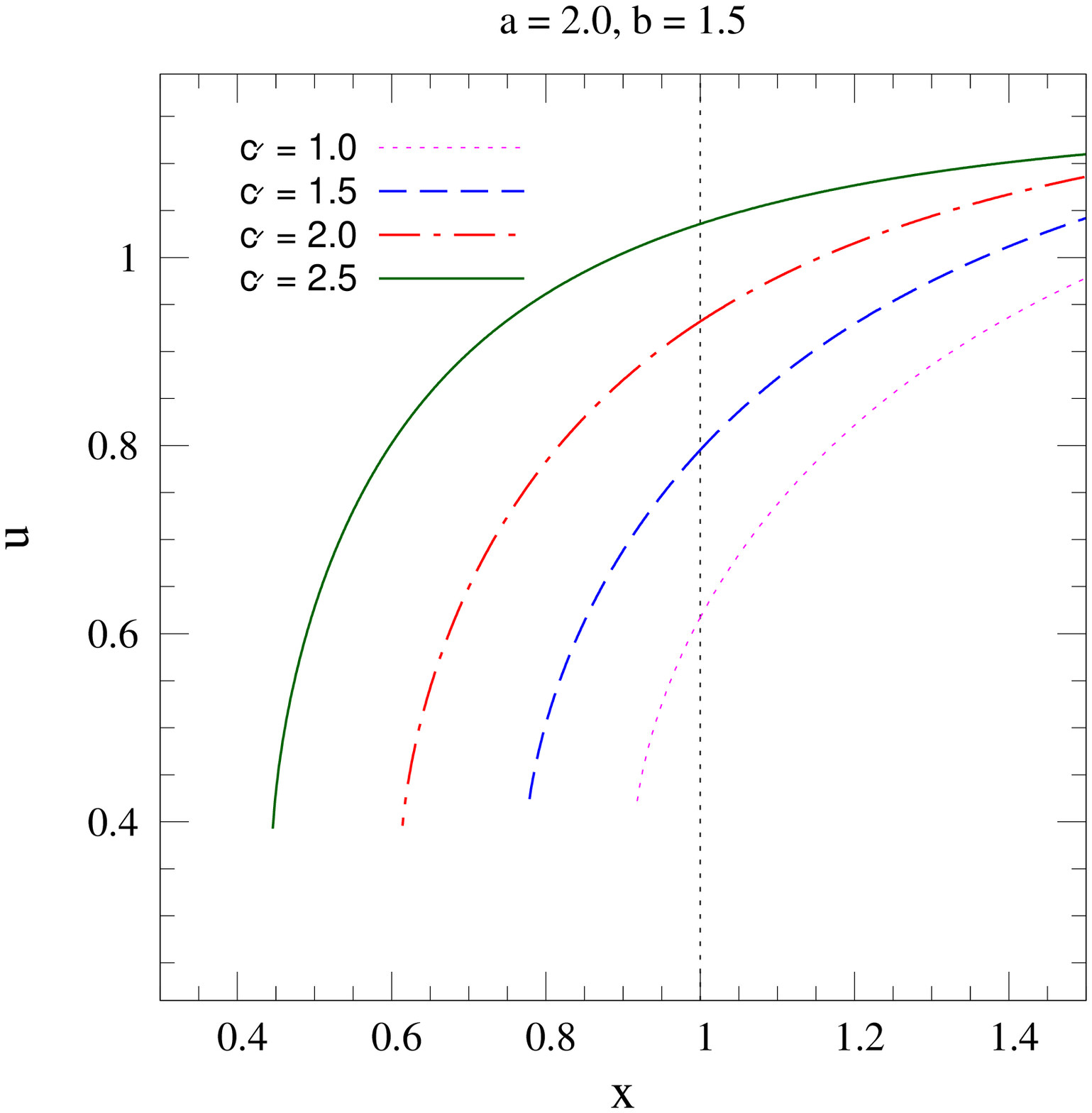}}
\label{fig_var_c}
}
\caption[]{The dependence of dimensionless velocity $u$ on the dimensionless
model parameters $a$, $b$ and $c'$ are shown in panels~(a), (b) and (c)
respectively. Values of parameters that remained fixed are given on the top of
each panel. Vertical dotted lines in each panel represent the critical
point $x=1$.
}
\label{fig_var}
\end{figure*}

We should also mention in passing that the magnetic field need not be
coherent on the galactic scales, for the CRs to transfer their energy
and momentum to the thermal gas. However one does need some degree of
coherence. First, for Alfv\'en waves to be resonantly amplified, 
the field should be coherent over scales much larger than the 
CR gyro radii ($r_g$), which is of order $10^{12} \epsilon$(GeV) cm,
where $\epsilon$ is the CR energy in units of GeV (Kulsrud, 2005).
This scale is clearly much smaller than typical scales in the galactic
wind. Second, energy transfer via the waves occurs only if either the
forward or the backward travelling wave (with respect to the field
direction) dominates the wave spectrum (Skilling, 1975, Eq.~9). 
Note that the Alfv\'en waves basically travel along the field lines
as they damp out, and unless this damping length is smaller than the
coherence scale of the magnetic field, waves going in one direction
would meet oppositely travelling waves, and the net wave heating would
be negligible. For the physical conditions in the outflow solutions, we
find this damping length is generally smaller than a fraction
of a parsec (cf. Kulsrud, 2005, Chapter 12). Thus we need magnetic
field coherence scales to be larger than this value. We also note that
momentum transfer (without the corresponding energy transfer via waves)
imposes a milder constraint that the field coherence scale be only much 
larger than $r_g$. Note that radio observations of several edge on
galaxies do show significantly coherent radial fields emanating from
the galaxy (Chyzy et al., 2006; Heesen et al, 2009a, b; Krause, 2008).  

The main advantage of our work, is that one can predict a mass outflow
rate given other parameters. The restriction on the mass outflow rate 
arises because the solution is constrained to be regular at the wind
critical point. In this particular example of a $10^{12}~M_\odot$ galaxy, 
we find the mass outflow rate is $4\pi q = 16.2~M_\odot$~yr$^{-1}$.
This should be compared with the required star formation
rate of ${\dot M}_{SF} \sim (10/f) M_\odot {\rm yr}^{-1}$,
giving $\eta_w \sim 1.6 f \sim 0.16$ [for $f\sim 0.1$]. 

Note that the mass outflow rate is directly proportional to the value
of $C_e$ that one assumes. A value of $C_e$ larger by a factor $10$,
would also lead to $10$ times higher value of $q$, for the same value of $a$.
This can be directly seen from Eq.~\ref{eqna}. The value of $\eta_w$
however remains the same as both $q$ and ${\dot M}_{SF}$ increase by
the same factor. Our model~B is an example of such solution where we assume
$10$ times higher $4\pi C_e = 3 \times 10^{42}$ erg s$^{-1}$, keeping
all other parameters same as in model~A. Increasing $C_e$ by a factor
$10$ leads to $10$ times higher values of $P$, $P_c$ and $\rho$
and $B$ higher by a factor $\sqrt{10}$. 

\begin{table*}
\caption[]{The summary of various wind models.
 }
\begin{center}
\begin{tabular}{c c c c c c c c c c} 
\hline \hline
 & \multicolumn{8}{c}{Models} \\
\cline{2-10} \\
\raisebox{2ex}[0cm][0cm]{Parameters}  & A & B & C & D & E & F & G & H & I \\ 
\hline \hline
$a$  & 2.0 & 2.0 & 2.0 & 2.0 & 2.0 & 2.0 & 0.5 & 2.0 & 2.0 \\ \hline
$b$  & 1.5 & 1.5 & 1.5 & 1.5 & 1.5 & 1.5 & 1.5 & 4.0 & 1.5 \\ \hline
$c'$  & 1.5 & 1.5 & 1.5 & 1.5 & 1.5 & 1.5 & 1.5 & 1.5 & 2.5  \\ \hline
mass ($M_\odot$)  & $10^{12}$ & $10^{12}$ & $10^{11}$ & $10^{10}$ & $10^{9}$ & $10^{8}$ & $10^{9}$ & $10^{9}$ & $10^{9}$  \\ \hline
$r_{vir}$ (kpc)  & 100 & 100 & 30.7 & 11.9 & 4.7 & 1.7 & 4.7 & 4.7 & 4.7 \\ \hline
$l$ (kpc)$^{\dagger}$  & 10 & 10 & 3.1 & 1.19 & 0.47 & 0.17 & 0.47 & 0.47 & 0.78 \\ \hline
$c$  & 15 & 15 & 15 & 15 & 15 & 15 & 15 & 15 & 15 \\ \hline
$4\pi C_e$ (erg~s$^{-1}$)  & $3\times 10^{41}$ & $3\times 10^{42}$ & $3\times 10^{40}$ & $3\times 10^{39}$ & $3\times 10^{38}$ & $3\times 10^{37}$ & $3\times 10^{38}$ & $3\times 10^{38}$ & $3\times 10^{38}$ \\ \hline \hline
$r_0$ (kpc)  & 7.8 & 7.8 & 2.4 & 0.92 & 0.36 & 0.13 & 0.43 & 0.30 & 0.35 \\ \hline
$P$ (eV~cm$^{-3}$)  & 0.59 & 5.9 & 1.10 & 1.44 & 1.84 & 2.66 & 0.55 & 2.19 & 1.09  \\ \hline
$P_c$ (eV~cm$^{-3}$)  & 0.50 & 5.0 & 0.93 & 1.22 & 1.56 & 2.25 & 0.46 & 1.08 & 0.71 \\ \hline
$B$ ($\mu$G)  & 6.2 & 19.8 & 8.5 & 9.8 & 11.0 & 13.2 & 7.8 & 18.0 & 7.5 \\ \hline
$n$ (cm$^{-3}$)  & $2.7\times 10^{-3}$ & 0.027 & 0.015 & 0.077 & 0.388 & 2.03 & 0.097 & 0.388 & 0.231 \\ \hline
$T$ (K)  & $2.6 \times 10^6$ & $2.6 \times 10^6$ & $8.5 \times 10^5$ & $2.2 \times 10^5$ & $5.5 \times 10^4$ & $1.5 \times 10^4$ & $6.5 \times 10^4$ & $ 6.5 \times 10^4$ & $5.4 \times 10^4$ \\ \hline
$V_c$ (km~s$^{-1}$)$^{\ddagger}$  & 271 & 271 & 155 & 79 & 40 & 21 & 40 & 40 & 40  \\ \hline
$v_\infty$ (km~s$^{-1}$)  & 292 & 292 & 166 & 85 & 42 & 22  & 66 & 38 & 31 \\ \hline
$F/\sqrt{2}$ & 1.08 & 1.08 & 1.07 & 1.08 & 1.05 & 1.05  & 1.65 & 0.95 & 0.78 \\ \hline
$4\pi q$ ($M_\odot$yr$^{-1}$)  & 16.2 & 162 & 4.99 & 1.93 & 0.76 & 0.28 & 0.19 & 0.76 & 1.27 \\ \hline
$\eta_w^\S$  & 0.16 & 0.16 & 0.5 & 1.9 & 7.6 & 28 & 1.9 & 7.6 & 12.7 \\ \hline \hline
\multicolumn{9}{l}{$^\dagger$ $l=r_{vir}/10$} \\
\multicolumn{9}{l}{$^\ddagger$ calculated at $r=l $} \\
\multicolumn{9}{l}{$^\S$ $\eta_w$ is calculated taking $f=0.1$}
\end{tabular}
\label{table1}
\end{center}
\end{table*}

\subsection{Wind solutions for different galaxy masses}

We now consider wind solutions for a range of galaxy masses adopting
the same dimensionless parameters, $a=2.0$, $b=1.5$ and $c'=1.5$. 
For each galaxy mass we fix a virial radius assuming the galaxy
(and its dark matter halo), collapsed from a $3\sigma$ density
fluctuation in a standard LCDM cosmology. The critical point is
taken to be at $l=r_{vir}/10$. Further we assume that the energy
outflow rate $C_e$ is proportional to the galaxy mass, and is given by
$4 \pi C_e = 3 \times 10^{41} (M/10^{12} M_\odot)$ erg s$^{-1}$.
Such a scaling assumes that a fixed fraction of the galactic mass
is converted to stars over a fixed timescale. Thus the star formation
rate and hence the rate of energy input by SNe is proportional to the
mass of the galaxy (normalised as earlier to a SFR of
$\sim (10/f) M_\odot$ yr$^{-1}$ for a massive $10^{12} M_\odot$ galaxy).
The derived physical parameters of the wind solutions, for galaxy masses
ranging from $10^{11}-10^8~M_\odot$ are presented in Table~\ref{table1},
as models~C to F respectively. 

Several trends in the derived physical parameters can be seen from
Table~\ref{table1}. As the galaxy mass decreases from $10^{12}-10^{8}~M_\odot$,
the gas and cosmic ray pressures at the critical point increase by
a factor $\sim 4$, from about $0.5-2$~eV~cm$^{-3}$, while $B$ increases
by a factor $\sim 2$ from $6.2-13.2~\mu$G. There is a much larger change
in the density of the wind material, which increases by nearly a factor
$\sim 10^3$, from $\sim 0.003-2.1$~cm$^{-3}$, as the galaxy mass decreases
by 4 orders of magnitude. Correspondingly the temperature of the wind which
is proportional to $P/\rho$, decreases from about $2.6 \times 10^6$~K,
for a $10^{12}~M_\odot$ galaxy to $T\sim 1.5 \times 10^4$~K for a
$10^8~M_\odot$ galaxy. The increased density and decreasing temperature
associated with these winds for lower mass galaxies implies that gas
cooling would become more and more important for them. We will return
to this aspect below.

Further, for the models A to F in Table~\ref{table1}, we obtain very
similar values of $F$, with $F/\sqrt{2} \sim 1.05-1.08$. Therefore,
in all these cases the asymptotic wind speed follows closely the circular
velocity of the galaxy at the critical point, with the factor $F/\sqrt{2}$
of order unity. There can however be a factor of $2$ scatter in this relation
for the range of dimensionless parameters $a$, $b$ and $c'$ that we
consider (see models~E, G, H and I of Table~\ref{table1}). Moreover,
the mass loading factor follows very closely the scaling with circular
velocity at the critical point predicted by analytic arguments, 
$\eta_w/f \propto V_c^{-2}$. Thus while massive high redshift galaxies
have a mass loading factor of $\sim 0.2$, small mass galaxies, with total
mass $M \sim 10^9-10^8~M_\odot$ have a large $\eta_w \sim (7.6-28)$
for $f\sim 0.1$. In Samui, Subramanian \& Srianand (2008, 2009)
we found that the volume filling of the IGM by outflows is 
mainly dominated by small mass galaxies with $M\lesssim 10^9~M_\odot$. 
Hence the results on small mass galaxies obtained
here have very important implications for such models.

\begin{figure*}
\centerline{\includegraphics[width=12.0cm,angle=-90]{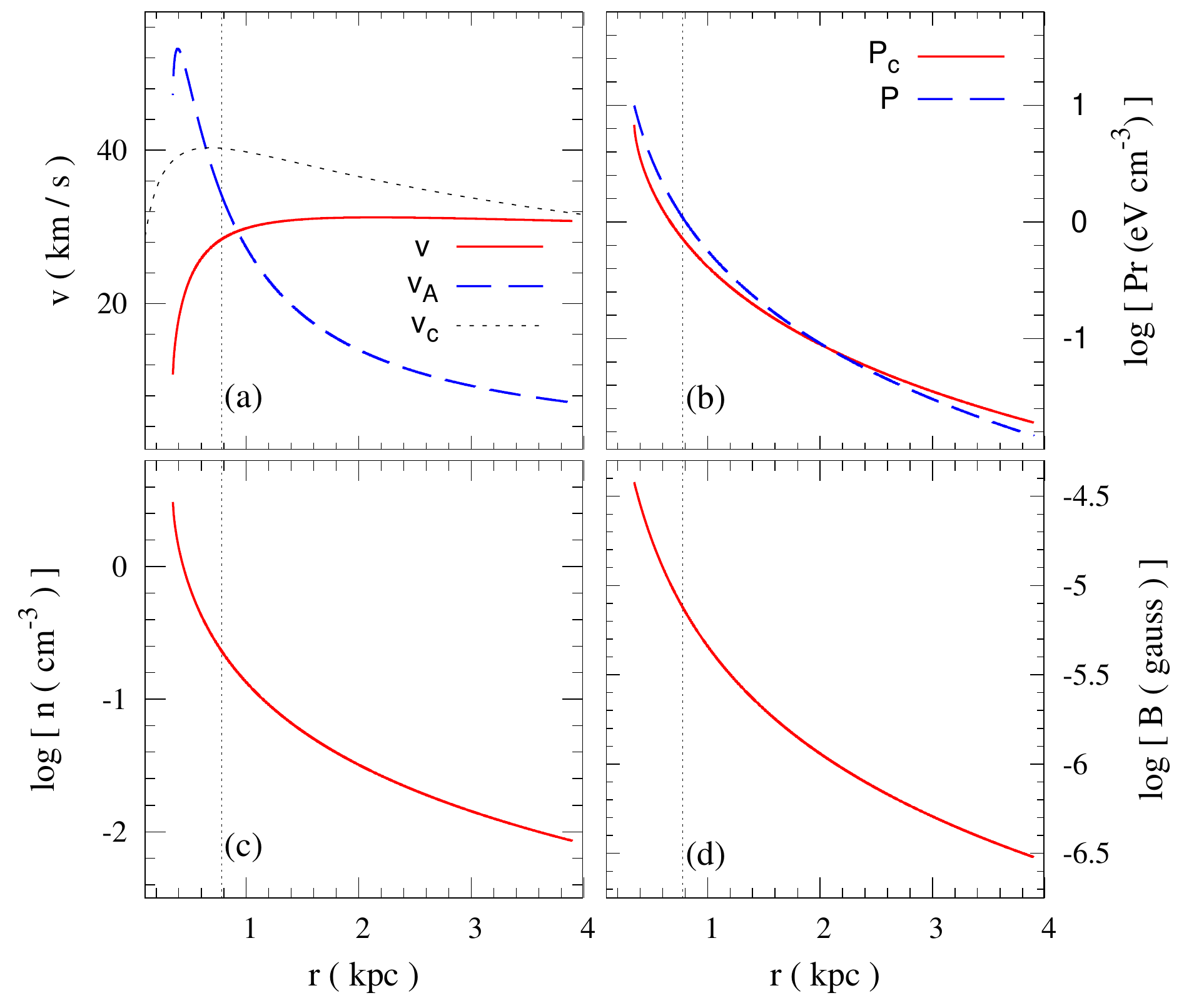}}
\caption[]{Characteristics of cosmic ray driven wind coming from a
$10^{9}~M_\odot$ galaxy. We show velocity, thermal gas pressure, cosmic ray
pressure, the density of the thermal gas and the magnetic field
as a function of radial distance from the galactic centre. The vertical
dotted lines in each panel represent the critical point.
}
\label{fig_wind_M9}
\end{figure*}

\subsection{Varying the dimensionless parameters}

Let us now consider the effect of dimensionless parameters $a$, $b$
and $c'$. To begin with, we concentrate on their impact on the
dimensionless velocity profile. In all transonic solutions $u$
starts from some finite value $u_0$ and at a finite $x=x_0$ and
increases till it passes through the critical point at $x=1$.
Note that the mass conservation equation, Eq.~\ref{eqn_21}, 
implies that the solutions have to start at finite $x=x_0$, and with
a finite velocity, to avoid infinite densities. This feature can be
changed if we have source terms in both the mass and energy conservation
equations at least upto $x=x_0$. One of our aims is also to look for
transonic solutions where $x_0$ is as small as possible, such that 
sources need not extend to the critical point $x=1$.

In Fig.~\ref{fig_var_a} we keep a fixed $b=1.5$ and $c'=1.5$ 
and vary $a$. The starting point $x_0$ decreases as one adopts 
larger and larger $a$, till about $a=2$, where it reaches
$x_0=0.78$. Further increase in $a$ does not lead to a decrease in
$x_0$. Also as can be seen from Fig.~\ref{fig_var_a}, $u_c$
increases with $a$, a feature which follows from Eq.~\ref{eqn_41},
which predicts $u_c \propto a^{1/2}$. Importantly, we find that transonic
solutions are not possible for arbitrarily small values of $a$.
For example one requires $a \gtrsim 0.5$ to have transonic solutions
adopting $b=1.5$ and $c'=1.5$. 

The effect on the velocity profile of varying $b$ and $c'$ are shown 
respectively in Fig.~\ref{fig_var_b} and Fig.~\ref{fig_var_c}. 
As one increases $b$ or $c'$ keeping other parameters fixed, $x_0$ decreases,
and the starting point for the wind can be at a smaller radius. Thus a
larger value for the magnetic field or a more concentrated galaxy
potential aids in starting the outflow at a smaller radius
\footnote{Such changes in $x_0$ also alters $f$, but from our work in
Appendix~\ref{appB}, by at most a factor of $\sim 1.3$.}.
At the same time, a minimum $b$ and $c'$ are required to get
a transonic solution, with other parameters fixed. 

We now consider the changes to the wind solutions in dimensional form,
which obtain on changing the adopted dimensionless parameters. Our models~G, H
and I study this aspect. We focus on the case when the galaxy mass
is $10^9~M_\odot$ because of the importance of such small mass galaxies to
the volume filling of the IGM. We find that decreasing the parameter
$a$ by a factor $4$ (i.e. model~G) basically decreases the mass outflow
rate by the same factor. This is as expected from Eq.~\ref{eqna},
where we see that $q \propto a$. The decrease in $a$ also brings
the base of the outflow, $r_0$, closer to the critical point $l$.
Model~H shows that increasing $b$ basically leads to a solution requiring
a larger magnetic field, but starts the flow at smaller radius compared to
the critical point.

Finally, we have illustrated the effect of increasing the parameter
$c'$ from $1.5$ to $2.5$ in model~I. In Fig.~\ref{fig_wind_M9} we have
shown the characteristics of this model. Since we wish to keep the NFW
potential form similar, we have adopted the same concentration parameter $c$, 
which implies that the critical point has to occur at a larger radius; 
$l/r_{vir} =0.17$ for this model instead of $0.1$. As also mentioned
earlier, we see that increasing $c'$ allows the wind to start at an
even smaller $r_0/l \sim 0.45$. It also leads to a larger mass outflow
rate, with $4\pi q = 1.28 M_\odot$ yr$^{-1}$, larger by a factor of $1.7$
compared to the solution with $c'=1.5$. The cosmic ray pressure again
begins slightly smaller than the gas pressure at the base of the outflow
$r=r_0=0.35$ kpc, with values of $P_c =0.7$~eV~cm$^{-3}$ and
$P=1.1$~eV~cm$^{-3}$ at the critical point $r=0.78$ kpc, but exceeds the
gas pressure asymptotically. The asymptotic wind velocity is $31$~km~s$^{-1}$,
and in this case smaller than the circular velocity at the critical point, but
comparable to the asymptotic circular velocity. The wind density is
$\sim 0.23$~cm$^{-3}$ at the critical point and then falls as $1/r^2$
asymptotically. The magnetic field strength is of order $7.5 \mu$G at
the critical point.

\subsection{Cooling of the wind material}

\begin{figure}
\centerline{
{\includegraphics[width=6.5cm,angle=-90]{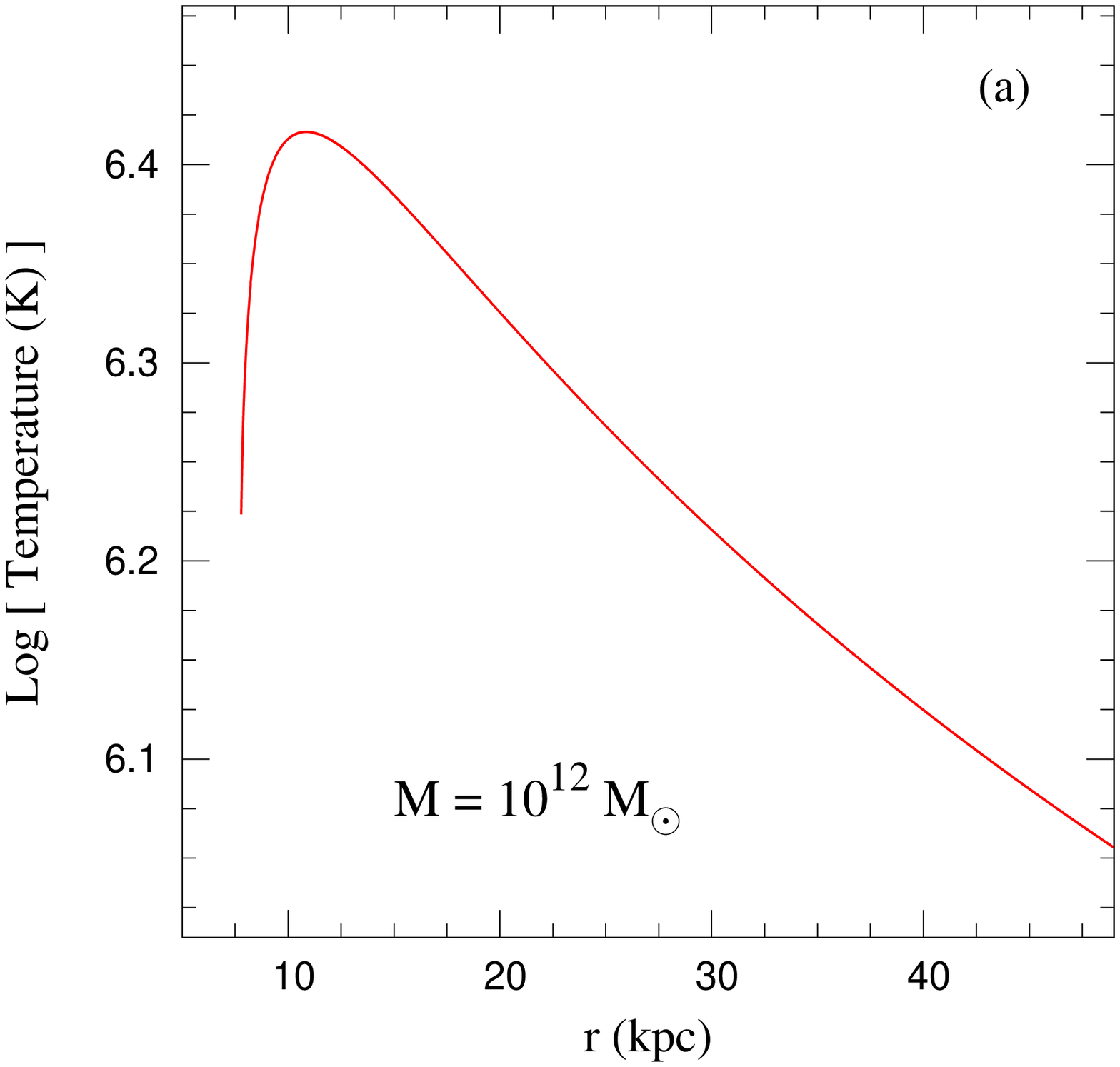}}
}
\centerline{
{\includegraphics[width=6.5cm,angle=-90]{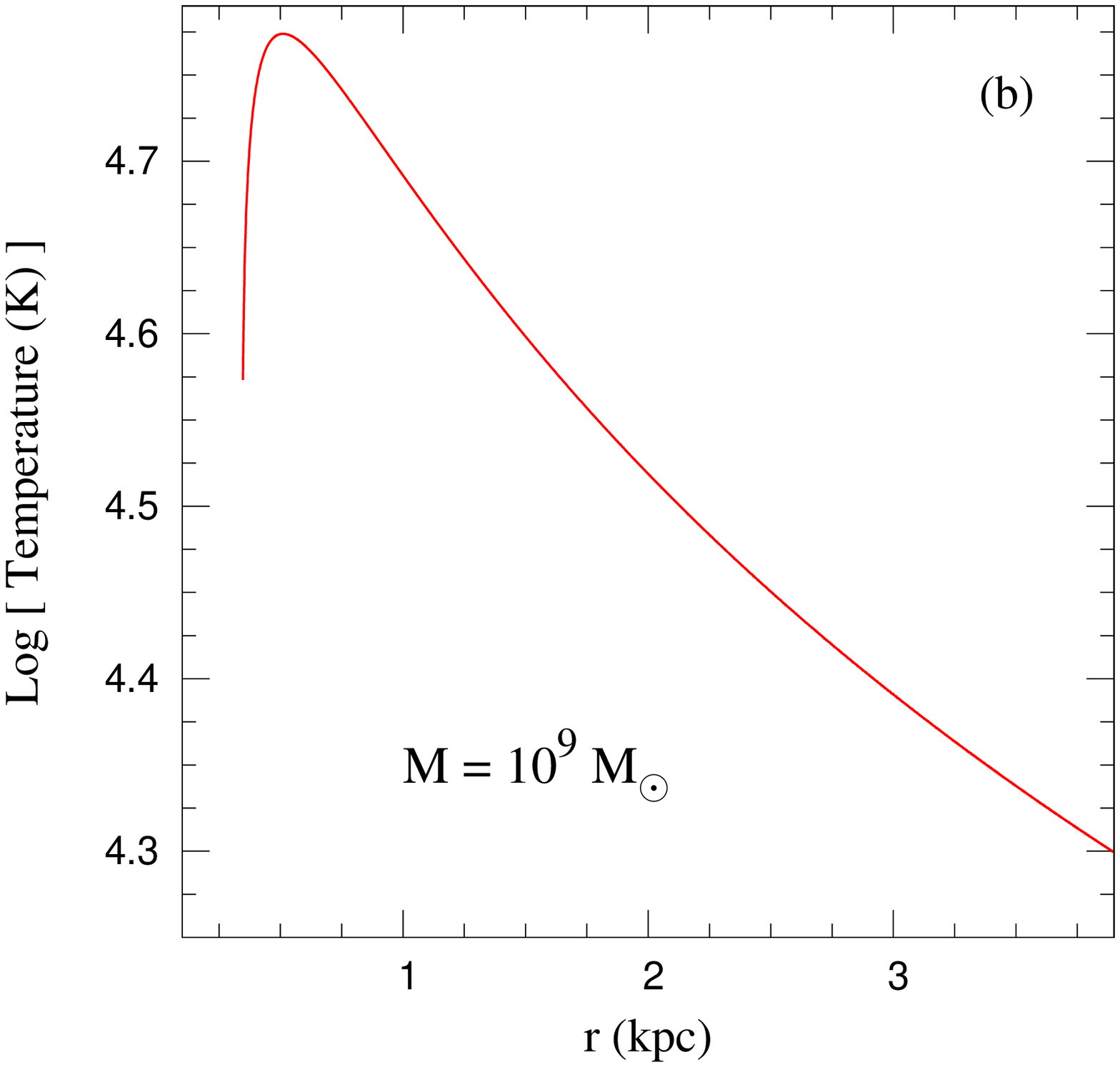}}
}
\caption[]{Temperature profile for the models~A and E
of Table~\ref{table1} are shown in panel~(a) and (b) respectively.}
\label{fig_temp}
\end{figure}

The wind solutions presented above do not take into account radiative cooling
of the thermal gas. Note that in these solutions the density of the wind
material at the critical point is $n(l) \sim 0.003-2$~cm$^{-3}$ for galaxy
masses in the range $10^{12}-10^{8}~M_\odot$. The corresponding temperatures
of the wind at the critical point are $2.6\times 10^6-1.5\times 10^4$~K for this
mass range of galaxies. We show the temperature profile for the
model~A and E of Table~\ref{table1} in top and bottom panels of
Fig.~\ref{fig_temp} respectively.
The radiative cooling of the gas depends on the temperature,
density as well as the metallicity of the gas. Taking a typical value for
the cooling function $\Lambda_N = 5\times 10^{-23}$~erg~cm$^3$~s$^{-1}$
(Sutherland \& Dopita, 1993), the cooling times are $t_c \sim 10^8$~yr and
$t_c \sim 10^3$~yr for $10^{12}~M_\odot$ and $10^{8}~M_\odot$ galaxies respectively.
This should be compared with the advective time scale of the wind at the
critical point, $t_a\sim l/v(l)$. We find $t_a \sim 5\times 10^7~$yr
and $10^8$~yr for the $10^{12}~M_\odot$ and $10^{8}~M_\odot$ galaxies
respectively. Therefore, radiative cooling is very important especially
for low mass galaxies. However the cosmic ray pressure by itself could
drive a wind even if the thermal gas cools rapidly, as was found by
Ipavich (1975), for outflows in a point mass potential.
We therefore examine in the next section, if one can
obtain similar cold wind solutions (where the thermal pressure
is neglected), from galaxies in a NFW halo potential.

\section {Cold wind solutions}

\subsection{Basic formalism}
We assume that radiative cooling is so efficient that most of the thermal
energy is lost from the gas, and so neglect the influence of the thermal
pressure compared to the CR pressure in the momentum equation. However, 
cosmic rays can still transfer momentum (via Alfv\'en waves) to the thermal gas. 
The equations of mass conservation (i.e. Eq.~\ref{eqn_con_q}), momentum
conservation (Eq.~\ref{eqn_2}) with the thermal pressure neglected,
and the equation governing cosmic ray evolution (Eq.~\ref{eqn_4}
and Eq.~\ref{eqn_I}) still hold. And they form by themselves a closed
system describing the evolution of $\rho, v$ and $ P_c$.
Again we assume that the magnetic field falls of as $1/r^2$.
From Eq.~\ref{eqn_con_q}, Eq.~\ref{eqn_4} and Eq.~\ref{eqn_I},
we still get Eq.~\ref{eqn_13} of Appendix~A. Substituting this in
Eq.~\ref{eqn_2}, neglecting $P$ compared to $P_c$, gives 
\begin{equation}
\rho v \frac{\de v}{\de r} + c_*'^2 \frac{\de \rho}{\de r} = -\frac{\rho GM(r)}{r^2}
\label{eqn_15_a}
\end{equation}
where now
\begin{equation}
c_*' = \left[\frac{4}{3} \frac{P_c}{\rho}
\frac{(v + v_A/2)}{(v+v_A)}\right]^{1/2}
\label{eqn_c_star_prime}
\end{equation}
is the effective sound speed. Eliminating $\de \rho/\de r$ using
Eq.~\ref{eqn_12} we obtain the cold wind equation
\begin{equation}
\frac{\de v}{\de r} = \frac{2v}{r}\frac{\left[1-\cfrac{GM(r)}{2rc_*'^2}\right]}
{\left[\cfrac{v^2}{c_*'^2}-1\right]}.
\label{eqn_dv_dr_cold}
\end{equation}
Hence, the cold wind equation (Eq.~\ref{eqn_dv_dr_cold}) has the same
form as the hot wind equation (Eq.~\ref{eqn_dv_dr}); only the effective
sound speed has now changed since we are neglecting the thermal pressure
of the gas. Thus the arguments relating the mass loading factor
and the circular velocity of the galaxy given in section 2, and leading
to  Eq.~\ref{eqn_eta}, are still valid, except that effective $f$ is
likely to be smaller due to radiative cooling.

In order to obtain numerical solutions, we again use the transformations
of Eq.~\ref{transform}, and get
\begin{equation}
\rho^* u x^2 = 2,
\label{eqn_47}
\end{equation}
\begin{equation}
2\frac{\de u}{\de x} + \frac{4a}{ux^2}
\left[\ln (1+c' x) - \frac{c' x}{1+c' x}\right] + 
x^2\frac{\de \theta_\Pi}{\de x} = 0,
\label{eqn_48}
\end{equation}
\begin{equation}
\frac{\de}{\de x}\left\{4\theta_{\Pi}x^2\left[u + \frac{(ub)^{1/2}}{x}\right]\right\}
= x^2\left[ u + \frac{(ub)^{1/2}}{x}\right] \frac{\de \theta_\Pi}{\de x}.
\label{eqn_cr_energy}
\end{equation}
Basically Eqs.~\ref{eqn_47} and \ref{eqn_48} come from Eqs.~\ref{eqn_con_q}
and \ref{eqn_2} and Eq.~\ref{eqn_cr_energy} comes from combining
Eqs.~\ref{eqn_4} and \ref{eqn_I}. Note that in this case $C_e$ used
in the transformation equations (Eq.~\ref{transform})
is just an arbitrary constant and does not represent the conserved
luminosity of the system like in Eq.~\ref{eqn_18}.

\begin{figure*}
\centerline{\includegraphics[width=12.0cm,angle=-90]{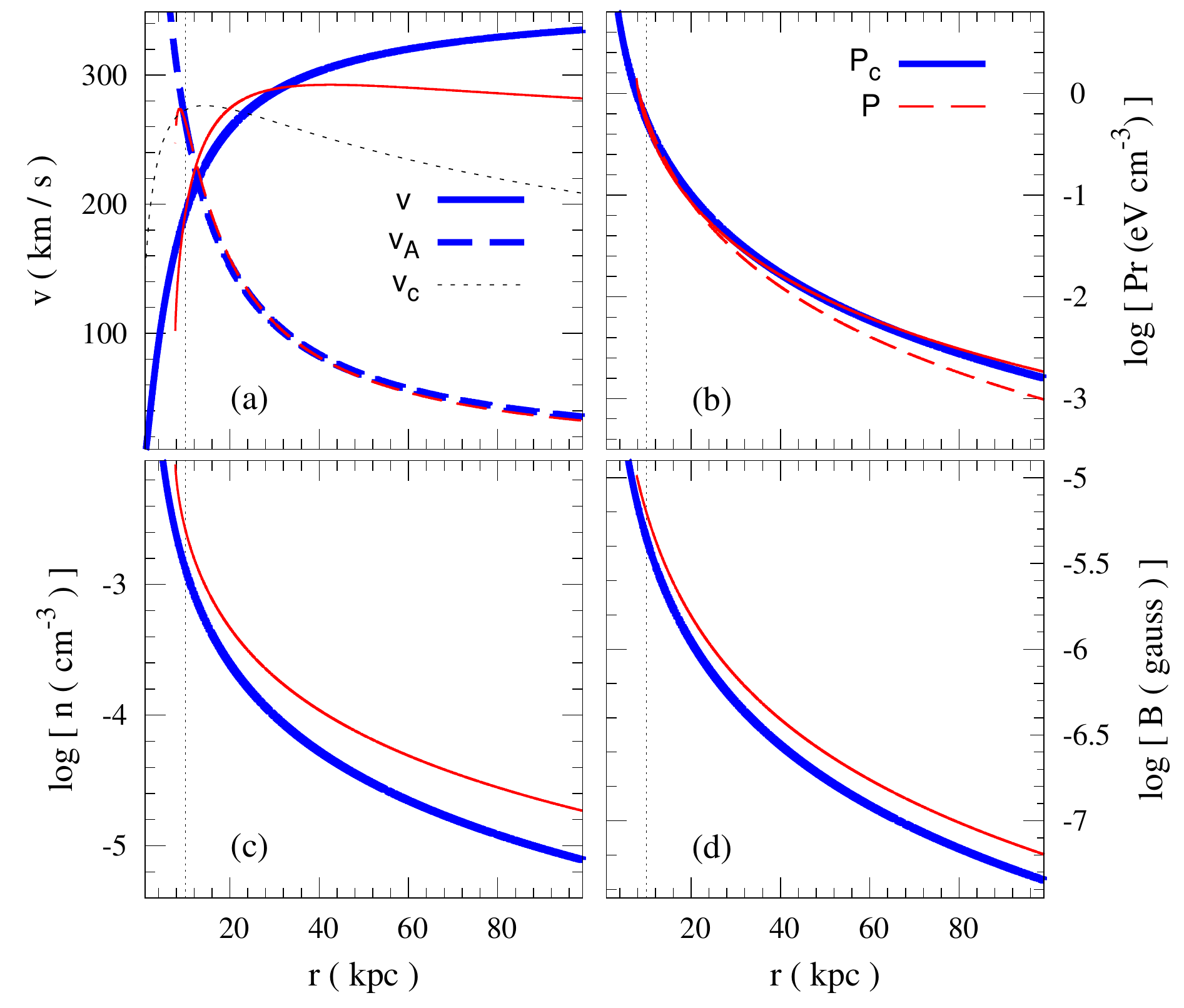}}
\caption[]{Characteristics of cosmic ray driven cold wind coming from a
$10^{12}~M_\odot$ galaxy (thick lines). We show velocity, thermal gas pressure,
cosmic ray pressure, the density of the thermal gas and the magnetic field
as a function of radial distance from the galactic centre. The vertical
dotted lines in each panel represent the critical point. For comparison we
also show the corresponding hot wind solution with thin lines.
}
\label{fig_wind_M12_cold_a}
\end{figure*}

Following the same procedure as in section 3, we arrive at two simultaneous
differential equations for the flow:
\begin{equation}
\frac{\de \theta_\Pi}{\de x} = \cfrac{- \cfrac{4}{3}\theta_\Pi\left[\cfrac{2}{x}
+ \cfrac{1}{u}\cfrac{\de u}{\de x}\right]\left[u + \cfrac{(ub)^{1/2}}{2x} \right]}
{\left[ u + \cfrac{(ub)^{1/2}}{x}\right]}
\label{eqn_50}
\end{equation}
and
\begin{equation}
\frac {\de u}{\de x} = \frac{2u}{x} \frac{\xi - \cfrac{a}{x}\left\{
\ln\left(1+c'x\right)-\cfrac{c'x}{1+c'x}\right\}}{u^2 - \xi},
\label{eqn_51}
\end{equation}
where
\begin{equation}
\xi = \frac{2}{3}\theta_\Pi u x^2\left[\frac{u+\cfrac{(ub)^{1/2}}{2x}}
{u+\cfrac{(ub)^{1/2}}{x}}\right].
\label{eqn_53}
\end{equation}
Note that Eq.~\ref{eqn_50} and Eq.~\ref{eqn_26} are identical 
(a feature also found by Ipavich, 1975).

The flow again has a critical point when both numerator and denominator
of Eq.~\ref{eqn_50} vanish and we again choose the critical point
to be at $x=1$. At the critical point,
\begin{equation}
u_c^2 = a\left[\ln \left(1 + c'\right) - \frac{c'}{1+c'}\right] = a {\mathfrak{F}}(c') = \xi_c
\label{eqn_54}
\end{equation}
and 
\begin{equation}
\theta_\Pi ^c = \frac{3}{2} u \left[\frac{u+(ub)^{1/2}}{u+\cfrac{(ub)^{1/2}}{2}}\right].
\label{eqn_pi_c_cold}
\end{equation}
Using the L'Hospital's rule, we evaluate the dimensionless velocity gradient
at the critical point,
\begin{equation}
{\cal A}\left(\frac{\de u}{\de x}\right)^2 + {\cal B}\left(\frac{\de u}{\de x}\right) + {\cal C} = 0
\label{eqn_57}
\end{equation}
where
\begin{eqnarray}
{\cal A} &=& 10 \beta^2 + 31\beta + 28,
\qquad {\cal B}= 4u(4 - 5\beta - 2\beta^2), \nonumber\\
{\cal C} &=& -4u^2(5\beta^2 + 14 \beta + 2) + \frac{12ac'^2}{(1+c')^2}(1+\beta)(2+\beta),
\nonumber \\
&&\beta = v_A/v = (ub)^{1/2}/u, \nonumber
\end{eqnarray}
with $u$ in Eqs.~\ref{eqn_pi_c_cold} and \ref{eqn_57} is evaluated at $x=1$.
We now turn to the numerical solution of the cold wind equations
for a range of galaxy masses.

\subsection{Cold wind solution from a massive galaxy}

\begin{figure*}
\centerline{\includegraphics[width=12.0cm,angle=-90]{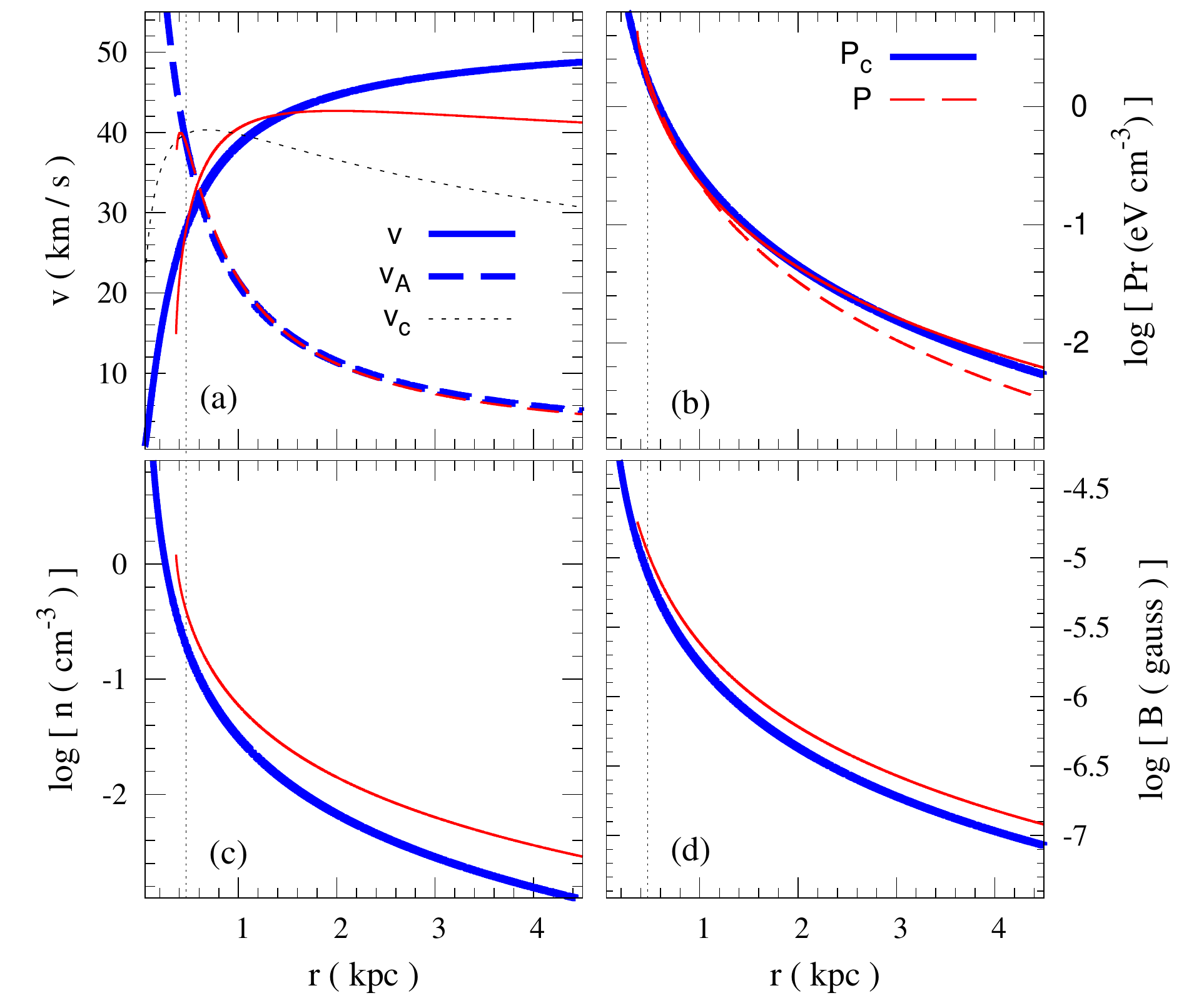}}
\caption[]{Characteristics of cosmic ray driven cold wind coming from a
$10^{9}~M_\odot$ galaxy (thick lines). We show velocity, thermal gas pressure,
cosmic ray pressure, the density of the thermal gas and the magnetic field
as a function of radial distance from the galactic centre. The vertical
dotted lines in each panel represent the critical point. For comparison we
also show the corresponding hot wind solution with thin lines.
}
\label{fig_wind_M9_cold_a}
\end{figure*}

Consider first CR driven cold wind from a massive, $10^{12}~M_\odot$ galaxy.
We choose all the parameters, except $C_e$, as in model~A. Recall that
the value of $C_e$ is an arbitrary constant for the cold wind solutions.
We fix its value by demanding that at the critical point, $r=l=10$ kpc,
the cosmic ray pressure is similar to that which obtains for the hot wind
solution described in the previous section (see model~A of Table~\ref{table1}).
We do this, so as to directly compare the mass outflow rate ($4\pi q$)
in the two cases. For this purpose, it turns out that one needs to adopt a
value for $4\pi C_e = 1.5\times 10^{41}$~erg~s$^{-1}$, half the value
used in the hot wind solution of Fig.~\ref{fig_wind_M12}. The resultant
dimensional cold wind solution is shown in Fig.~\ref{fig_wind_M12_cold_a}
as thick lines. We also show for comparison, the corresponding
hot wind solution given in Fig.~\ref{fig_wind_M12}, as thin lines.

The first feature to note (panel~(a)) is that, the cold wind solution
can start from an arbitrarily small radius and velocity; however from mass
conservation the density would then blow up. Having an upper limit on the
density will then decide the location of the base. Such a feature is also
seen in the cold wind solutions obtained by Ipavich (1975). This is unlike
the hot wind solution, where the flow starts from a finite radius with
velocity significant fraction of the local circular velocity. The final
asymptotic velocity of the cold wind solution turns out to be higher than
the corresponding hot flow, but still close to the circular velocity at
the critical point, with $F/\sqrt{2}= 1.26$. The cosmic ray pressure
through out the flow is similar to that of previous solution. However,
both the density and the magnetic field are smaller in case of cold wind
solution. The mass out flow rate for this solution is
$4\pi q = 8.1~M_\odot$~yr$^{-1}$, half the value which obtains for the
hot wind solution. Therefore the cosmic ray pressure can by itself
drive an outflow, with significant mass loss, even if the gas efficiently
radiates away its thermal energy. Note that were we to lower the adopted
value of the parameter $C_e$, keeping all other parameters the same,
one would also get a valid solution, with lower values for
$P_c, \rho \propto {C_e}$ and $B \propto \sqrt{C_e}$. The velocity profile
will however not change. Lowering $C_e$ would also imply a lower value of
$q$, with $q\propto C_e$.

\subsection{Cold wind solution for small galaxies}

We now turn to the small mass galaxies. In Fig.~\ref{fig_wind_M9_cold_a}
we show the cold wind solution (by thick lines) for a galaxy of mass
$10^9~M_\odot$, adopting the parameters as in model~E except for $C_e$.
We adjust $C_e$ so that the cosmic ray pressure for the cold and
hot wind solution of Fig.~\ref{fig_wind_M9}, are similar at the critical point.
This hot wind solution is also shown in Fig.~\ref{fig_wind_M9_cold_a}
as thin lines, for comparing with the cold wind solution.
From the figure, it is clear that the qualitative features of the 
cold wind solution are similar for a small mass galaxy and a massive galaxy.
The flow again starts with a small velocity near the centre of the galaxy
potential. The final velocity is larger than that of hot wind solution but
still related to the circular velocity with $F/\sqrt{2} = 1.25$.
The mass outflow rate is $4\pi q=0.38~M_\odot$~yr$^{-1}$ and is
again a factor 2 lower for the cold wind compared to the hot outflow.
However the exact reduction in the mass outflow rate again
depends on the parameters of the solution adopted as
discussed above for the case of winds from massive galaxies.

Recall that the mass loading factor scaled as $\eta_w \propto V_c^{-2}$
for the hot wind solutions without radiative cooling.
We have seen that the mass outflow rate for the cold wind solutions
scales with the mass outflow rate for the hot wind.
Thus we expect that the scaling relation between the mass loading factor
$\eta_w$ and the circular velocity, $\eta_w \propto V_c^{-2}$, will
continue to hold for the cold wind solutions, albeit with a 
smaller absolute value of $q$, for any particular galaxy mass.
In order to make a more quantitative statement, one needs to
solve for wind solutions with cooling included in the energy equation.
This is beyond the scope of this exploratory work.

\section{Discussion and Conclusions}

We have studied here cosmic ray driven wind solutions relevant
to galaxies whose gravitational potential is of the NFW form.
Cosmic rays are expected to be copiously generated in high
redshift star forming galaxies, (and also in present day starbursts), 
where the SFR and hence the SNe rate, is likely to be much higher
than those in our Galaxy. The CR driven winds extensively studied in the
context of our Galaxy, could then also be relevant for such
galaxies. Moreover, transonic wind solutions, with the fluid
velocity being accelerated from subsonic to supersonic speeds
exist for a $\gamma=5/3$ thermal gas, only if one can have energy
and/or momentum input till the critical point.
A purely thermally driven wind solution assuming such energy input 
upto the critical point is given by Chevalier and Clegg (1985).
On the other hand, cosmic rays naturally provide 
such energy and/or momentum input,
even if the energy sources (i.e. SNe) are well within the critical point,
or the thermal energy is lost due to radiative cooling.
These considerations formed the key motivation for our work. 

We have first given a suite of such galactic free wind solutions,
driven by both thermal and cosmic ray pressures, in an NFW gravitational
potential, where the physical parameters of the wind are fairly
reasonable. The solutions start at radii smaller than the critical point,
with $r_0/l \sim 0.4-0.8$, depending upon the dimensionless parameters.
For galaxy masses in the range $10^{8}-10^{12}~M_\odot$, the gas and
cosmic ray pressures at the critical point range respectively from
about $2-0.5$ eV cm$^{-3}$, while $B$ decreases from $13.2-6.2~\mu$G.
These values are comparable to the cosmic ray energy density of our Galaxy 
and magnetic field strengths in nearby spirals. The density of the
wind material at $r=l$, which ranges from $\sim 2.1-0.003$~cm$^{-3}$,
for the same range in galaxy masses, is also comparable to
the densities expected in galaxies. Correspondingly the temperature
of the wind increases from about
$T\sim 1.5 \times 10^4$K for a $10^8~M_\odot$ galaxy to
$2.6 \times 10^6$ K, for a $10^{12}~M_\odot$ galaxy.
The increased density and decreasing temperature
associated with these winds for lower mass galaxies
imply that gas cooling would become
more important for such galaxies.

It appears that $\mu$G strength magnetic fields are required
for CRs to drive a transonic wind in the NFW potential that
we have adopted. We do find that it is possible to find
cold wind solutions with very small magnetic fields in a point
mass potential. This issue is of interest as 
Zweibel (2003) has shown that even fields several orders of
magnitude smaller than galactic fields can accelerate and
confine CRs. The magnetic field that is required, need not however be
coherent over large scales, as we discussed earlier. 
Radio observation of several edge on galaxies do show significant
radial fields emanating from the galaxy, and associated with outflows
(Heesen et al, 2009a, b; Krause, 2008).  

All these solutions importantly have an asymptotic
wind velocity closely related to the circular velocity
of the galaxy. Such a correlation seems to be indeed observed
in outflows of ultraluminous infrared galaxies (Martin, 2005 Fig. 7).
Martin (2005) also finds that wind velocity is proportional to
the SFR of the galaxy. In our models, larger mass galaxies have
a larger SFR, and also a larger circular velocity and hence wind velocity.
Thus one naturally expects to find such a correlation in the CR driven
wind models.  

Our wind solutions also predict an anti-correlation between the
mass loading factor of the wind and the circular velocity, with 
$\eta_w/f \propto V_c^{-2}\propto v_{cir}^{-2}$.
We find that larger mass galaxies with $M\sim 10^{11}-10^{12}~M_\odot$,
typically have $\eta_w \sim 0.5-0.2$, adopting $f=0.1$.
Such mass loading factors are consistent with $0.1\le \eta_w \le 0.7$ determined
by Rupke, Veilleux \& Sanders (2002) for 
ultraluminous infrared galaxies at redshifts $0.04 \le z \le 0.27$.   
For smaller mass galaxies with total mass $M \sim 10^{10}-10^9~M_\odot$,
we obtain a larger  $\eta_w \sim 2-7.6$, while for galaxies 
with $M\sim 10^8~M_\odot$ this increases to $\eta_w \sim 28$.

Note that in our wind models the wind temperature varies
between $0.6-2.2 \times 10^5$ K, for dwarf galaxies with
total mass in the range $10^9-10^{10} M_\odot$.
The wind can have larger
temperatures for smaller values of the dimensionless parameter `a'
(by a factor upto 4) and smaller values of the parameter `l'
(the location of the critical point).
Thus galactic free winds in our models can have temperatures 
required to produce X-ray emission only in massive galaxies but not in
dwarf galaxies. However, several local dwarf galaxies  
do show extended X-ray emission indicative of hot gas, with
temperature of order KeV (cf. Martin, 1999; Ott et al., 2005), 
even though starbursting dwarf galaxies 
are only a small fraction (about $6\%$) of local low-mass population
(Lee et. al. 2009). 
If this X-ray emission is related to the wind material then
this hot component of the wind cannot be explained by the
models discussed here, where the source region is confined to
be within the sonic point.

If one does wish to construct very hot free wind solutions in such 
dwarf galaxies, one will have to use a model like that of
Chevalier and Clegg (1985), where the sources do extend
at least upto the radius where the wind velocity becomes sonic. 
Moreover, for such solutions, the radiative losses cannot be
more than about 30\% (cf. Silich, Tenorio-Tagle and
Rodriguez-Gonzalez, 2004). It may be that both
these conditions are obtained for some dwarf galaxies.
But they need not hold in general; i.e. the sources need not in general
extend to the sonic point and the physical conditions could
be such that radiative cooling is important. In such cases, our work
shows that Cosmic rays could still help to drive a wind.

Indeed, in our cold wind solutions, we consider models where,
due to radiative cooling, the thermal pressure can be neglected
compared to the cosmic ray pressure. We find that even under
these circumstances, cosmic rays can nevertheless drive a healthy
transonic wind. The flow starts at a small radius with a small
velocity which become supersonic beyond a critical point,
reaching an asymptotic value of order the circular velocity
of the galaxy. We show that in such solutions, and depending
on the model parameters, the mass outflow rates are smaller,
by at least a factor 2, compared to the hot CR driven wind solutions.
Nevertheless, even in this case, we still expect that smaller mass
galaxies have a higher $\eta_w$. A large $\eta_w$ for small mass
galaxies imply that cosmic ray driven outflows could provide a
strong negative feedback to the star formation in dwarf galaxies.

We have assumed spherical symmetry in our wind
solutions. It will be useful to find such solutions in
disk like geometry as well. 
We have also not included the effect of CR diffusion.
Such diffusion could lead to a reduction of the
mass outflow rate as shown by Dorfi (2004).
Further, we have assumed the wind to be free, and 
not confined by any external pressure, say due to gas in 
the halo of the galaxy. In case such confining agents are important,
one could first have a stage of filled hot bubble evolution,
as in models described in our earlier work 
on thermally driven outflows into the halo/IGM 
(Samui, Subramanian \& Srianand, 2008). And the free wind
solution would become relevant in latter stages,
when the confining medium has been evacuated out.

The results on small mass galaxies obtained
here, that their winds have asymptotic speeds close to their
circular velocity, and have a large mass loading factor, would
have important implications for models of metal
enrichment of the IGM.
Several groups have shown that outflows from such small mass galaxies are
possibly the most important contributors to the volume filling
of the IGM with metals (cf. Madau, Ferrara and Rees, 2001; 
Samui, Subramanian \& Srianand, 2008).
Thus it is important to revisit such global models
in the light of the results obtained in this work. 

\section*{acknowledgements}
We thank an anonymous referee for several useful comments.

\appendix

\section{The wind equation}
\label{appA}

Using $\rho v r^2=q$ in Eq.~\ref{eqn_3}, we obtain,
\begin{equation}
\rho v \left[ v \frac{\de v}{\de r} + \frac{5}{2}\frac{\de}{\de r}
\left(\frac{P}{\rho}\right)\right] = - \rho v \frac{GM(r)}{r^2} - \left(v+v_A\right)
\frac{\de P_c}{\de r}.
\label{eqn_6}
\end{equation}
Multiplying Eq.~\ref{eqn_2} by $v$, and subtracting from Eq.~\ref{eqn_6},
we get,
\begin{equation}
v \left[-\frac{\de P}{\de r} - \frac{\de P_c}{\de r} + \frac{5\rho}{2}
\frac{\de}{\de r}
\left(\frac{P}{\rho}\right)\right] = - \left(v+v_A\right)
\frac{\de P_c}{\de r}
\label{eqn_7}
\end{equation}
which can be simplified to give 
\begin{equation}
\frac{\de P}{\de r} - \frac{5}{3}\frac{P}{\rho}\frac{\de \rho}{\de r} =
- \frac{2}{3}\frac{v_A}{v}\frac{\de P_c}{\de r}.
\label{eqn_8}
\end{equation}
The Alfv\'en velocity is given by, 
\begin{equation}
v_A = \frac{B}{\sqrt{4\pi \rho}}
\label{eqn_9}
\end{equation}
where we assume that the magnetic field satisfies the condition
\begin{equation}
Br^2 = {\rm constant},
\label{eqn_10}
\end{equation}
which leads to
\begin{equation}
\frac{1}{v_A}\frac{\de v_A}{\de r} = -\frac{2}{r} - \frac{1}{2\rho}
\frac{\de \rho}{\de r}.
\label{eqn_11}
\end{equation}
Also Eq.~\ref{eqn_con_q} can be differentiated to get
\begin{equation}
\frac{1}{\rho}\frac{\de \rho}{\de r} + \frac{1}{v}\frac{\de v}{\de r} + \frac{2}{r} = 0.
\label{eqn_12}
\end{equation}
Hence from Eq.~\ref{eqn_4}, Eq.~\ref{eqn_12} and Eq.~\ref{eqn_11} we get 
the cosmic ray pressure gradient as,
\begin{equation}
\frac{\de P_c}{\de r} = \frac{4}{3} \frac{P_c}{(v + v_A)}\left[v + \frac{v_A}{2}\right]
\frac{1}{\rho}\frac{\de \rho}{\de r}.
\label{eqn_13}
\end{equation}
The gas pressure gradient can be obtained using Eq.~\ref{eqn_8} and Eq.~\ref{eqn_13},
\begin{equation}
\frac{\de P}{\de r} = \left[\frac{5}{3}\frac{P}{\rho}-\frac{8}{9}\frac{P_c}{\rho}
\frac{(v+v_A/2)}{(v+v_A)}\frac{v_A}{v}\right]\frac{\de \rho}{\de r}.
\label{eqn_14}
\end{equation}
Substituting Eq.~\ref{eqn_13} and Eq.~\ref{eqn_14} 
in the momentum equation (Eq.~\ref{eqn_2}), we get,
\begin{equation}
\rho v \frac{\de v}{\de r} + c_*^2 \frac{\de \rho}{\de r} = -\frac{\rho GM(r)}{r^2}
\label{eqn_15}
\end{equation}
where the effective sound speed $c_*$ is determined by,
\begin{equation}
c_*^2 = \frac{5}{3}\frac{P}{\rho} + \frac{4}{3}
\frac{P_c}{\rho}
\frac{(v + v_A/2)(3v-2v_A)}{3v(v+v_A)}.
\label{eqn_c_starA}
\end{equation}
Eliminating $d\rho/dr$ using Eq.~\ref{eqn_12},
we get the wind equation given in the main text,
\begin{equation}
\frac{\de v}{\de r} = \frac{2v}{r}\frac{\left[1-\cfrac{GM(r)}{2rc_*^2}\right]}
{\left[\cfrac{v^2}{c_*^2}-1\right]}.
\label{eqn_dv_dr2}
\end{equation}

\section{Hot wind with uniform source distribution}
\label{appB}

We consider here the hot wind solution in presence of energy sources
upto certain radius $R$. This will allow us to estimate what fraction
of total SNe energy ends up as the kinetic energy of the wind material.
We assume that upto $r=R$, SNe explosions provide energy to the thermal
gas and cosmic rays as well. Let ${\cal L}_{th}$ and ${\cal L}_{cr}$
are the luminosity per unit volume injected to the thermal gas and
cosmic rays respectively. We take both ${\cal L}_{th}$ and ${\cal L}_{cr}$
as constant for $r< R$ and zero out side. These two source terms have
to be added to the right hand side the energy equation (Eq.~\ref{eqn_3})
and the cosmic ray evolution equation (Eq.~\ref{eqn_4}). Note that
$ (4/3)\pi R^3 ({\cal L}_{th} + {\cal L}_{cr}) = (4/3)\pi R^3
{\cal L}_0 = L_0$, where $L_0$ is the total luminosity that is being
pumped by the SNe to the wind material. Apart from energy, SNe explosions
would also add mass and momentum to the system. Hence
the modified fluid and cosmic rays equations can be written as
(compared to Eqs.~\ref{eqn_con_q}-\ref{eqn_4}),
\begin{equation}
\frac{1}{r^2}\frac{\de}{\de r}(\rho v r^2) = \left\{
\begin{array}{ll}  Q = {\rm constant} & \textrm{if $r<R$} \\
0 & \textrm{if $r \ge R$} \end{array}\right.
\label{eqn_con_q_source}
\end{equation}
\begin{equation}
\rho v \frac{\de v}{\de r} + \frac{\de P}{\de r} + \frac{\de P_c}{\de r} +
\frac{GM(r)\rho}{r^2} = \left\{
\begin{array}{ll}  -Qv & \textrm{if $r<R$} \\
0 & \textrm{if $r \ge R$} \end{array}\right.
\label{eqn_2_source}
\end{equation}
\begin{eqnarray}
&&\frac{1}{r^2}\frac{\de}{\de r}\left[\rho v r^2 \left(\frac{1}{2}v^2 + 
\frac{5}{2}\frac{P}{\rho} \right)\right] \nonumber \\
&& \hskip 1cm = \left\{
\begin{array}{ll} - \rho v r^2 \cfrac{GM(r)}{r^4} + I + {\cal L}_{th} & \textrm{if $r<R$} \\
- \rho v r^2 \cfrac{GM(r)}{r^4} + I & \textrm{if $r \ge R$} \end{array}\right.
\label{eqn_3_source}
\end{eqnarray}
\begin{equation}
\frac{1}{r^2}\frac{\de}{\de r}\left[4P_c r^2 \left(v + v_A\right)\right]=
\left\{ \begin{array}{ll}  -I + {\cal L}_{cr} & \textrm{if $r<R$} \\
-I & \textrm{if $r \ge R$} \end{array}\right.
\label{eqn_4_source},
\end{equation}
where $Q$ is the mass input rate per unit volume to the system.
Further, Eq.~\ref{eqn_I} for the energy exchange between cosmic rays and
thermal gas remains the same.

One can integrate Eq.~\ref{eqn_con_q_source} to get
\begin{equation}
\rho v r^2 = \left\{
\begin{array}{ll}
\cfrac{Qr^3}{3}& \textrm{if $r<R$} \\
q & \textrm{if $r \ge R$}. \end{array}\right.
\label{eqn_int_q}
\end{equation}
Note that, $q$ is the same constant as in Eq.~\ref{eqn_con_q}.
Demanding the continuity of $\rho$ and $v$ across $r=R$, we obtain,
\begin{equation}
q = \cfrac{QR^3}{3}.
\label{eqn_Q_q}
\end{equation}

We can also add Eqs.~\ref{eqn_3_source} and \ref{eqn_4_source}
and integrate the resulting equation to get an algebraic conservation law.
For $r\ge R$ the source free equations are still valid and we will get
the conservation law, Eq.~\ref{eqn_18}. However, for $r<R$, we get,
\begin{eqnarray}
& &\hskip -0.7cm\rho v r^2\left[ \frac{1}{2}v^2 + \frac{5}{2}\frac{P}{\rho} \right]
+ \frac{QGM}{3 {\mathfrak{F}}(c)}\left(\frac{r_{vir}}{c}\right)^2\frac{1}{4}
\left[2\left\{\left(\frac{cr}{r_{vir}}\right)^2-3\right\}
\right.\times \nonumber \\ 
& & \hskip 1.5cm \left. \ln\left(1+
\frac{cr}{r_{vir}}\right) - 3\left\{\left(\frac{cr}{r_{vir}}\right)^2
-2\frac{cr}{r_{vir}}\right\}\right]
\nonumber \\ 
& & \hskip 2.5cm + 4P_c r^2 \left(v + v_A\right) =
\frac{{\cal L}_0r^3}{3}.
\label{eqn_18_source}
\end{eqnarray}

Again demanding continuity across $r=R$, from Eqs.~\ref{eqn_18} and
\ref{eqn_18_source} (and also using the relation of Eq.~\ref{eqn_Q_q}), we obtain,
\begin{eqnarray}
& & \hskip -0.7cm 4\pi C_e = L_0 - 
 \frac{3\pi qGM}{{\mathfrak{F}}(c)R}\left[
2\left\{1-\left(\frac{r_{vir}}{cR}\right)^2\right\}\ln\left(1+
\frac{cR}{r_{vir}}\right) \right. \nonumber \\
& & \hskip 3cm \left. + \left(2\frac{r_{vir}}{cR} - 1\right)\right].
\label{eqn_ce_L0}
\end{eqnarray}
The second term in the right hand side comes from the gravitational
term only. Note that $4\pi C_e$ is the total energy flux that is passing
through any sphere and it is also the final wind kinetic energy. So,
the final kinetic luminosity of the wind is the total luminosity that is
being supplied by the SNe minus the energy per unit time required to climb
out of the gravitational potential of the galaxy. Hence the fraction $f$ of
Eq.~\ref{eqn_M_dot_w_L0} entirely depends on the gravitational potential
of the galaxy if radiative cooling is negligible. If the total luminosity
is less than the required potential energy, there would not be any outflowing
solution. This threshold luminosity would depend on the shape
of the gravitational potential and the distribution of the
energy sources with respect to that potential.

In order to determine $f$ for NFW galaxy potential and the uniform source
distribution, we assume that $L_0$ is higher than the threshold
value and hence $4\pi C_e$ is positive. We use Eqs.~\ref{eqn_M_dot_w_L0}
and \ref{eqn_M_dot_w_q} to rewrite Eq.~\ref{eqn_ce_L0} and find
\begin{equation}
f = \cfrac{1}{\left[1 + \cfrac{3}{2}\cfrac{c}{{\mathfrak{F}}(c)(F/\sqrt{2})^2}
\cfrac{v_{cir}^2}{V_{c}^2}
{\mathcal F}(\zeta)\right]},
\label{eqn_f}
\end{equation}
where 
\begin{equation}
{\mathcal F} (\zeta) = 2\zeta\left(1-\zeta^2\right)\ln\left(
1+\cfrac{1}{\zeta}\right)+\zeta\left(2\zeta-1\right)
\label{func_F}
\end{equation}
with
\begin{equation}
\zeta = \frac{r_{vir}}{cR}.
\nonumber
\end{equation}
Note that $v_{cir} \approx V_c$ for a nearly flat rotational
curve.

\begin{figure}
\centerline{\includegraphics[width=6cm,angle=-90]{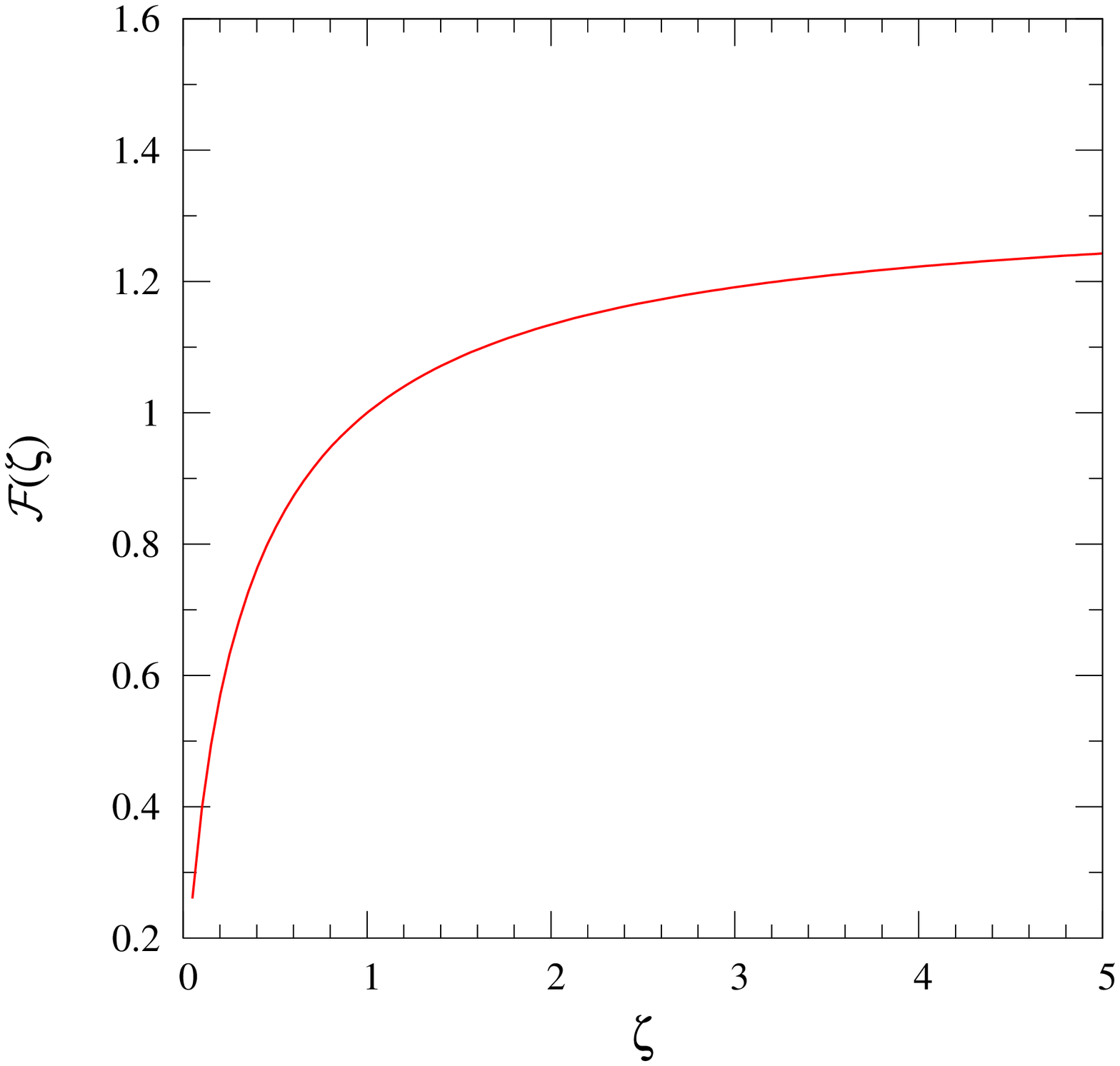}}
\caption[]{The function $ {\mathcal F}$ of Eq.~\ref{func_F}
is shown as a function of $\zeta$. It reaches to an asymptotic value
of 1.3 as $\zeta\to \infty$.}
\label{fig_F}
\end{figure}

The function ${\mathcal F}$ is shown in Fig.~\ref{fig_F}.
It is clear that ${\mathcal F}$ is of order unity for a wide range
of $\zeta$ values. For example, if we take $R=r_{vir}/12$ and $c=15$,
we get $\zeta = 0.8$ and then ${\mathcal F}=0.94$. Further,
${\mathfrak{F}}(c) = 1.84 $ for $c=15$. We will see from numerical
solution that $F/\sqrt{2}\approx 1$. Putting these
values in Eq.~\ref{eqn_f}, we obtain $f= 0.08$. Hence most of the
total luminosity has been used to climb out of the gravitational potential
and only about 10\% ends up as wind kinetic energy.

\end{document}